\documentclass[11pt]{article}
\usepackage{a4wide}
\newcommand{\be}{\begin{equation}}
\newcommand{\ee}{\end{equation}}
\newcommand{\ba}{\begin{array}}
\newcommand{\ea}{\end{array}}
\newcommand{\bea}{\begin{eqnarray}}
\newcommand{\eea}{\end{eqnarray}}
\newcommand{\X}{{\bf r},t}

\def\X{\scriptstyle\rm X}

\def\half{\frac{1}{2}}

\def\refnote#1{{$^{\hbox{\scriptsize #1}}$}}
\def\tablenote#1{\setbox0=\hbox{$^{\hbox{\scriptsize
#1}}$}\noindent\hangindent=\wd0 \box0}
\begin{document}
\title{\bf Excitation Energies from Time-Dependent Density Functional
Theory Using Exact and Approximate Potentials}
\author{M.~Petersilka and E.~K.~U.~Gross\\
{\sl Institut f\"ur Theoretische Physik, Universit\"at W\"urzburg,}\\ 
{\sl Am Hubland, 97074 W\"urzburg, Germany}\\ \\
Kieron Burke\\
{\sl Department of Chemistry, Rutgers University}\\
{\sl 610 Taylor Road, Piscataway, NJ 08854}}
\maketitle
\begin{abstract}
The role of the exchange-correlation potential and
the exchange-correlation kernel in the calculation of 
excitation energies from time-dependent density functional theory
is studied.
Excitation energies of the helium and beryllium atoms
are calculated, both from the exact Kohn-Sham ground-state potential,
and from two orbital-dependent approximations.
These are exact exchange and self-interaction corrected local
density approximation (SIC-LDA), both calculated using 
Krieger-Li-Iafrate approximation.
For the exchange-correlation kernels, three adiabatic approximations
were tested:  the local density approximation, exact exchange, and 
SIC-LDA.
The choice of the ground-state exchange correlation potential has the largest
impact on the absolute position of most excitation
energies.  In particular,
orbital-dependent approximate potentials
result in a uniform shift of the transition energies
to the Rydberg states.
\end{abstract}
\section{Introduction}
\large\sf

The Hohenberg-Kohn theorem
\cite{HohenbergKohn:64} of ground-state density functional theory
(DFT) guarantees that
every observable of a stationary physical system can 
be expressed in terms of its ground-state density.
In principle, this is also true for the set of excited-state
energies, and several extensions of ground-state DFT have been
proposed
\cite{GunnarssonLundqvist:76}-
\nocite{ZieglerRaukBaerends:77,vonBarth:79,Theophilou:79,
Kohn:86,GrossOliveiraKohn:88a,GrossOliveiraKohn:88b,OliveiraGrossKohn:88,
Nagy:90,Nagy:95,Levy:95}
\cite{Goerling:96}. 
Accurate calculations of excitation energies, however,
remain a difficult subject. 
Recently, some of us proposed a different approach to the calculation of
excitation energies \cite{PetersilkaGossmannGross:96},
within the framework of time-dependent DFT (TDDFT)
\cite{RungeGross:84}.
The central idea is to use the fact that the linear density response has
poles at the physical excitation energies and can be calculated 
from the response function of a noninteracting Kohn-Sham (KS) system
and a frequency-dependent Kohn-Sham (KS) kernel.
In this way, we obtain the shifts of the KS orbital differences (which
are the poles of the KS response function) towards the true excitation
energies. 
Recent applications 
\cite{PetersilkaGross:96}-
\nocite{BauernschmittAhlrichs:96,C96,BauernschmittEtAl:97,
Grabo:97,GisbergenEtAl:98,CCS98,CJCS98,GraboPetersilkaGross:99}
\cite{Sb99}
suggest that this method may become a standard tool in
quantum chemistry.

The success of any density functional method, however, depends
on the quality of the functionals employed.
In this article, we 
investigate the relative
importance of the approximations inherent in the TDDFT formalism
for the calculation of 
discrete excitation energies of finite systems.
This mainly concerns the role of the ground-state XC
potential, $v_{\rm xc}({\bf r})$ compared to
the dynamical XC kernel, $f_{\rm xc}({\bf r},{\bf r'};\omega)$.
For the helium and beryllium atoms, we compare the 
results obtained from using the exact XC potentials and two orbital
dependent potentials, one based on the exact exchange expression and
the other on the self-interaction corrected local density
approximation \cite{PerdewZunger:81}, evaluated with the method of Krieger,
Li and Iafrate
(KLI)
\cite{KriegerLiIafrate:90a}-
\nocite{KriegerLiIafrate:92b,KriegerLiIafrate:92c,
KriegerLiIafrate:95}
\cite{ChenEtAl:96} in combination with 
three distinct approximations for the XC kernels, which are
given in  
Sect.~\ref{sect:approx_for_fxc}.

\section{Formalism}

\subsection{Kohn-Sham equations for the frequency-dependent linear
density response}

The frequency dependent 
linear density response $n_{1\sigma}({\bf r},\omega)$ 
of electrons with spin $\sigma$, reacting to a 
perturbation $v_{1 \sigma'}$ of frequency $\omega$ can be written 
in terms of 
the interacting density-density response function
$\chi_{\sigma \sigma'}$ by
\cite{FetterWalecka:71} 
\begin{equation} \label{conventional}
n_{1\,\sigma}({\bf r},\omega)=
\sum_{\sigma'} \int\!d^{3}r'\,
\chi_{\sigma \sigma'}({\bf r},{\bf r'};\omega) \, 
v_{1 \, \sigma'}({\bf r'},\omega) \,.
\end{equation}
In the spin-dependent 
version \cite{LiuVosko:89} of time-dependent DFT
\cite{RungeGross:84}, the density response $n_{1 \sigma}$
can be expressed in terms of the response function 
$\chi_{s\, \sigma \sigma'}$ of the non-interacting Kohn-Sham (KS)
system \cite{PetersilkaGossmannGross:96,GrossDobsonPetersilka:96}:
\begin{equation} \label{dftalt}
n_{1\,\sigma}({\bf r},\omega)=
\sum_{\sigma'} \int\!d^{3}r'\,
\chi_{{\rm s}\,\sigma \sigma'}({\bf r},{\bf r'};\omega) \, 
v_{{\rm s},1 \, \sigma'}({\bf r'},\omega) \,.
\end{equation}
The KS response function
\begin{equation} \label{chi01}
\chi_{{\rm s}\, \sigma \sigma'}({\bf r},{\bf r}';\omega) =
\delta_{\sigma \sigma'} \sum_{j, k} (f_{k\sigma} - f_{j\sigma})
\frac{\varphi_{j\sigma}({\bf r}) \varphi_{k\sigma}^*({\bf r})
      \varphi_{j\sigma}^*({\bf r}') \varphi_{k\sigma}({\bf r}')}
     {\omega - (\epsilon_{j\sigma} - \epsilon_{k\sigma}) + i\eta} 
\end{equation} 
is readily expressed in terms of the 
unperturbed static Kohn-Sham orbitals $\varphi_{k\sigma}$
(with occupation numbers $f_{j \sigma}$).
Relation (\ref{dftalt})
contains the linearized KS potential 
\begin{equation} \label{vs1}
  v_{{\rm s,1}\,\sigma}({\bf r},\omega) = v_{{1}\sigma}({\bf r},\omega)
  + \sum_{\sigma'} \int \! d^3r' \, 
  \frac{n_{1\sigma'}({\bf r'},\omega)}{|{\bf r} - {\bf r'}|}
  +\sum_{\sigma'} \int \! d^3r' \, 
  f_{{\rm xc}\,\sigma\sigma'}({\bf r},{\bf r'};\omega)
  \, n_{1\sigma'}({\bf r'},\omega) \,. 
\end{equation}
in which
the spin-dependent exchange-correlation (XC) kernel $f_{\rm xc}$
is defined as the Fourier transform of
\begin{equation}
\label{fxc}
f_{{\rm xc}\,\sigma\sigma'}[n_{0 \uparrow}, n_{0 \downarrow}]
({\bf r},t,{\bf r'},t') :=
\left.
\frac{\delta v_{{\rm xc} \,\sigma }[n_{\uparrow},n_{\downarrow}]({\bf r},t)}
     {\delta n_{\sigma'}({\bf r'},t')}
\right|_{n_{0 \uparrow}, n_{0 \downarrow}} \,.
\end{equation}
Given an approximation to $f_{\rm xc}$, Eqs.~(\ref{dftalt}) and (\ref{vs1})
can be solved self-consistently for every frequency $\omega$.

\subsection{Approximations for the exchange-correlation
kernel}
\label{sect:approx_for_fxc}

For spin-unpolarized ground states, there are
only two independent combinations of the spin components of the XC
kernel, since 
$f_{{\rm xc} \,\uparrow \uparrow} = f_{{\rm xc} \,\downarrow \downarrow}$
and 
$f_{{\rm xc} \,\uparrow \downarrow} =
f_{{\rm xc} \,\downarrow \uparrow}$:
\be
f_{\rm xc}=\frac{1}{4}\sum_{\sigma\sigma'} f_{{\rm xc} \,\sigma
\sigma'}=\half (f_{{\rm xc} \,\uparrow \uparrow}+
f_{{\rm xc} \,\uparrow \downarrow})~~~~~~~
G_{\rm xc}= \frac{1}{4}\sum_{\sigma\sigma'} \sigma\sigma'f_{{\rm xc} \,\sigma
\sigma'}=\half
(f_{{\rm xc} \,\uparrow \uparrow}-
f_{{\rm xc} \,\uparrow \downarrow}),
\label{fxcGxc}
\ee
(contrary to common usage, we have not separated the Bohr magneton in the
definition of $G_{\rm xc}$).
Note that $f_{\rm x}=G_{\rm x}$, as exchange
contains only parallel spin contributions.

The simplest possible approximation is
the adiabatic
local (spin-)density approximation \cite{GrossKohn:90} (ALDA).
For spin-unpolarized
ground states, this leads to
\be \label{fxcALDA}
f_{\rm xc}^{\rm ALDA}[n]({\bf r},{\bf r'}) =
\delta({\bf r}-{\bf r'}) \,\left. \frac{d^2
e_{\rm xc}^{\rm hom}
}{d \rho^2} 
\right|_{\rho=n({\bf r}),\zeta=0}, ~~~~~~~
G_{\rm xc}^{\rm ALDA}[n]({\bf r},{\bf r'}) = 
\delta({\bf r}-{\bf r'}) 
\frac{\alpha_{\rm xc}(n(\bf r))}{n(\bf r)},
\ee
where $e_{\rm xc}^{\rm hom}$ is the 
exchange-correlation energy per unit volume of the homogeneous electron
gas,
$\zeta$ is the relative spin polarization, $(n_\uparrow-n_\downarrow)/n$,
and the spin-stiffness 
$
\alpha_{\rm xc} = 
\left.
\frac{\delta^2}{\delta \zeta^2}
\left(e_{\rm xc}^{\rm hom}(\rho,\zeta)/\rho\right)
\right|_{\zeta=0}
$.

Approximate XC functionals derived from the homogeneous electron gas 
suffer from several shortcomings, such as
spurious self-interaction contributions.
These are very significant for calculations of orbital eigenvalues,
as they affect the asymptotic decay of the ground-state potential.
For example, the XC potential in the local density approximation 
decays exponentially, so rapidly that only one virtual state is bound.
An alternative 
approach towards the construction of improved functionals is 
to use perturbation theory in the electron-electron coupling
constant\cite{GL93}.  This leads to orbital-dependent functionals,
which can be solved self-consistently using the
optimized effective potential method (OEP)
\cite{SharpHorton:53}-
\nocite{TalmanShadwick:76}
\cite{UllrichGossmannGross:95a}.
In the time-dependent case, this method
takes as a starting point a given (approximate) expression for the 
quantum mechanical action integral as a functional of a set of
orbitals\cite{L98}.
Variation with respect to a local effective potential then leads
to an integral equation for the exchange-correlation potential.
Given an exchange-correlation potential of that kind,
the corresponding
exchange-correlation kernel can be constructed in the same spirit
\cite{PetersilkaGossmannGross:96,PetersilkaGossmannGross:98}.
The essential steps are formally identical to the OEP construction of the
exchange-correlation potential for the ground-state\cite{GKKG98}.

In the time-dependent X-only approximation, $A_{\rm xc}$ is 
replaced by\footnote{In general,
 a Keldysh contour integral in complex time is needed
\cite{L98} to avoid causality difficulties\cite{GDP96}, except when the
action is local in the orbitals
in time, as is the case with all approximations tested here.}
\begin{equation}\label{opm1.3}
A_{\rm x-only} = -(1/2)\sum_{\sigma}\sum_{i,j}^{N_{\sigma}}
\int_{-\infty}^{t_{1}}\! dt\! \int\! d^{3}r\! \int\! d^{3}r'\:
\phi_{i\sigma}^{*}({\bf r}'t)\phi_{j\sigma}({\bf r}'t)\phi_{i\sigma}({\bf
r}t)
\phi_{j\sigma}^{*}({\bf r}t)\:/\:|{\bf r}-{\bf r}'|\,.
\end{equation}
The orbital-dependent exchange kernel in the time-dependent
KLI
approximation
is
\cite{PetersilkaGossmannGross:96,PetersilkaGossmannGross:98}
\begin{equation}
  \label{fxcoep}
  f_{{\rm x-only}\,\sigma\sigma'}^{\rm{TDOEP}}({\mathbf r},{\mathbf
r'}) = 
  - \delta_{\sigma\!\sigma'} {1\over|{\mathbf r}-{\mathbf r'}|}
  {\big|\sum_k 
    f_{k\sigma}\,\varphi_{k\sigma}({\mathbf r})\varphi_{k\sigma}^*({\mathbf
r'})\big|^2
    \over n_\sigma({\mathbf r})n_\sigma({\mathbf r'})}\;.
\end{equation}
In general, the exact $\X$-only kernel carries a frequency-dependence. 
This is not accounted for in the present approximation 
(\ref{fxcoep}). However, for one- and spin-unpolarized two-electron systems,
Eqs.~(\ref{fxcoep}) is 
the {\em exact} solution of the respective
integral equations in the limit of a time-dependent X-only theory.
This yields
\be
f_{\rm x}=G_{\rm x}=-\frac{2\big|\sum_k f_k \varphi_k({\bf r})
\varphi_k^*({\bf r'}) \big|^2}
{n({\bf r}) |{\bf r}-{\bf r'}| n({\bf r'})}~~~~~~~~
=\left( -\frac{1}{2|{\bf r}-{\bf r}'|}
{\rm ~~for~~2~~elec}\right).
\ee

Inherent to any X-only theory, the resulting kernels are lacking  
off-diagonal elements in spin space.
To improve upon the X-only treatment,
we use the self-interaction corrected
(SIC) LDA \cite{PerdewZunger:81} for 
$A_{\rm xc}$:
\begin{equation} \label{AxcSIC}
A_{\rm xc}^{\rm SIC} =
\int_{-\infty}^{t_1}\!\! \!dt
\left( E_{\rm xc}^{\rm LDA}[n_{\uparrow}(t),n_{\downarrow}(t)] 
  - \!\sum_{i \sigma}E_{\rm xc}^{\rm LDA}[n_{i \sigma}(t),0]
  - \frac{1}{2} \sum_{i \sigma} 
    \int\! d^3r \int\! d^3r'
    \frac{n_{i \sigma}({\bf r},t)\,n_{i \sigma}({\bf r'},t)}
         {|{\bf r}-{\bf r'}|}
\right)
\end{equation}
which is an orbital-dependent functional as well 
due to the explicit dependence on the orbital densities 
\begin{equation}
n_{i \sigma}({\bf r},t) = | \phi_{i \sigma}({\bf r},t) |^2  
\quad (i=1,2,\ldots,N/2) \,.
\end{equation}
]
An improvement over both ALDA and exact exchange might be
provided by correcting ALDA for self-interaction 
error\cite{PerdewZunger:81}.
Within the adiabatic SIC-LDA, the exchange-correlation kernel reads
\bea \label{fxc_oepsic}
\lefteqn{
f_{{\rm xc}\,\sigma \sigma'}^{\rm TDOEP-SIC}({\bf r},{\bf r'},\omega) =
f_{{\rm xc}\, \sigma \sigma'}^{\rm ALDA}({\bf r},{\bf r'},\omega) -}
\nonumber \\
& & {}- \frac{\delta_{\sigma \sigma'}}
     {n_{0\,\sigma}({\bf r}) \, n_{0\,\sigma}({\bf r'})}
\sum_{k} f_{k \sigma}
| \varphi_{k \sigma}({\bf r}) |^2 \, | \varphi_{k \sigma}({\bf r'}) |^2
\left( 
\frac{\delta v_{{\rm xc}\, \sigma}^{\rm LDA}(n_{k\,\sigma}({\bf
    r}),0)}
     {\delta n_{k\,\sigma}({\bf r'})}
+
\frac{1}{|{\bf r}-{\bf r'}|}
\right) \,.
\eea
This expression reduces to the exact result (Eq. (\ref{fxcoep}))
for one electron.  For more than one electron, spurious
self-interaction parallel-spin contributions in ALDA are corrected,
for both exchange and correlation.  The correction has no affect
on anti-parallel spin contributions, leaving simply the ALDA result.
We find simply
\be
f_{\rm xc}^{\rm SIC}=f_{\rm xc}^{\rm ALDA}+\Delta f_{\rm xc}^{\rm SIC}
~~~~~~
G_{\rm xc}^{\rm SIC}=G_{\rm xc}^{\rm ALDA}+\Delta f_{\rm xc}^{\rm SIC}
\ee
where
\be
\Delta f_{\rm xc}^{\rm SIC}=-\frac{2\sum_k f_k n_k({\bf r}) n_k({\bf r'})}
{n({\bf r}) n({\bf r}')} \left\{\delta({\bf r}-{\bf r}')\
\frac{\partial v^{\rm hom}_{\rm xc,\uparrow} (n_k, 0)}{\partial n_k ({\bf r})}
+\frac{1}{|{\bf r}-{\bf r}'|} \right\}.
\ee

\section{Calculation of excitation energies}

The linear density response has poles at the exact excitation 
energies of the interacting system (see, e.g., 
\cite{FetterWalecka:71}).
The key idea is to start from a particular KS  orbital
energy difference $\epsilon_{j \sigma} - \epsilon_{k \sigma}$
(at which the Kohn-Sham response function (\ref{chi01}) has a pole)
and to use the formally exact 
representation (\ref{dftalt}) of the linear density response 
to calculate the shifts of 
the Kohn-Sham excitation energies
 towards the true excitation energies $\Omega$.
To extract these shifts from the density response,
we cast Eq.~(\ref{dftalt}) together with (\ref{vs1})
into the form of an integral equation for $n_{1 \, \sigma}$: 
\bea
\label{Operatorgleichung}
\sum_{\nu '} \int \!\! d^3 y' & & \hspace{-0.5em}{}
\Bigg[
\delta_{\sigma \nu '} \delta({\bf r} - {\bf y'})
- \sum_{\nu} \int \!\! d^3 y 
\chi_{{\rm s}\, \sigma \nu} ({\bf r},{\bf y};\omega)
\Bigg(
\frac{1}{|{\bf y}-{\bf y'}|}
 + f_{{\rm xc} \,\nu \nu '}({\bf y},{\bf y'};\omega)
\Bigg) \Bigg]
n_{1 \nu '}({\bf y '},\omega)
                           \nonumber\\
& & {}=\sum_{\nu} \int d^3 y \,
\chi_{{\rm s}\,\sigma \nu} ({\bf r},{\bf y};\omega)
v_{1 \nu}({\bf y},\omega) \,.\label{aq12}
\eea
In general, the true excitation energies $\Omega$ are not
identical with the Kohn-Sham excitation energies $\epsilon_{j \sigma}
- \epsilon_{k \sigma}$, and the
right-hand side of Eq.~(\ref{Operatorgleichung}) remains finite for $\omega
\rightarrow \Omega$.
The exact spin-density response $n_{1 \sigma}$,
on the other hand, exhibits poles at the true excitation energies $\Omega$.
Hence,
the integral operator acting on $n_{1 \sigma}$ on the
left-hand side of Eq.~(\ref{Operatorgleichung}) cannot be invertible
for $\omega \rightarrow \Omega$.
This means that the integral operator acting on
the spin-density vector in Eq.~(\ref{Operatorgleichung}) is
non-invertible (i.e., has vanishing eigenvalues) at the 
physical excitation energies.
Rigorously,
the true excitation energies $\Omega$ are those
frequencies where the eigenvalues
$\lambda(\omega)$ of
\bea
\sum_{\nu '} \int d^3 y' \,
\sum_{\nu}   \int d^3 y  & &{} \hspace{-0.5em}
\chi_{{\rm s}\,\sigma \nu} ({\bf r},{\bf y};\omega)
\left(
\frac{1}{|{\bf y}-{\bf y'}|}
+ f_{{\rm xc}\, \nu \nu '}({\bf y},{\bf y'};\omega)
\right)
\gamma_{\nu '}({\bf y '},\omega) = \nonumber \\
& &{}= \lambda(\omega)
\gamma_{\sigma}({\bf r},\omega)
\label{ewgl}
\eea
satisfy
\begin{equation} \label{lambda=1}
\lambda(\Omega) = 1 \,.
\end{equation}

For notational brevity, we use double indices $q\equiv(j,k)$ to 
characterize an excitation energy;
$\omega_{q \sigma}\equiv\epsilon_{j \sigma}-\epsilon_{k
\sigma}$ denotes the excitation energy of the single-particle
transition $(k \sigma \rightarrow j \sigma)$.
Consequently, we set 
$ \alpha_{q \sigma} := f_{k \sigma}-f_{j \sigma} $ and
\begin{equation}
\label{bigphi}
\Phi_{q \sigma}({\bf r})
:=
\varphi_{k \sigma}^{*}({\bf r})\varphi_{j \sigma}({\bf r}) 
\end{equation}
as well as
\begin{equation}
\label{ZETA}
\xi_{q\sigma}(\omega)
:=
\sum_{\nu '} \int d^3 y' \,
\int d^3 y  \,
\Phi_{q \sigma}({\bf y})^*
\left(
\frac{1}{|{\bf y}-{\bf y'}|}
+ f_{{\rm xc}\, \sigma \nu '}({\bf y},{\bf y'};\omega)
\right)
\gamma_{\nu '}({\bf y '},\omega) \,.
\end{equation}
Without any approximation, 
equation (\ref{ewgl}) can be cast \cite{PetersilkaGross:96} into 
matrix form
\begin{equation}
  \label{matrixproblem}
  \sum_{\sigma'} \sum_{q'}
  \frac{M_{q \sigma \, q' \sigma'}(\omega)}
  {\omega - \omega_{q' \sigma'} + i \eta}
  \xi_{q' \sigma'}(\omega) = \lambda(\omega) \xi_{q \sigma}(\omega) \,,
\end{equation}
with the matrix elements
\begin{equation}
\label{excit9}
M_{q \sigma \, q' \sigma'}(\omega) =
\alpha_{q' \sigma'}
\int\!d^3 r \int\!d^3 r' \,
\Phi^{*}_{q \sigma}({\bf r})
\left(
\frac{1}{|{\bf r}-{\bf r'}|}
+ f_{{\rm xc} \,\sigma \sigma'}({\bf r},{\bf r'};\omega)
\right)
\Phi_{q' \sigma'}({\bf r'})  \,.
\end{equation}
At the frequencies $\omega = \Omega$,
Eq.\ (\ref{matrixproblem}) can be written as
\begin{equation} \label{eigenwertgl}
\sum_{q' \sigma'}
\left( M_{q \sigma \, q' \sigma'}(\Omega)
+ \delta_{q \sigma \, q' \sigma'} \omega_{q \sigma} \right)
\beta_{q' \sigma'}(\Omega) =
\Omega \beta_{q \sigma}(\Omega) \,,
\end{equation}
where we have defined
\begin{equation}
\beta_{q \sigma}(\Omega)
:= \xi_{q \sigma}(\Omega) / (\Omega - \omega_{q \sigma}) \,.
\end{equation}
The solutions $\Omega$ 
of the nonlinear matrix-equation (\ref{eigenwertgl})
are the physical excitation energies.
The inevitable truncation of
the infinite-dimensional matrix in Eq.\ (\ref{eigenwertgl}) amounts to
the approximation of $\chi^{(0)}$ by a finite sum
\begin{equation}
\chi^{(0)}({\bf r},{\bf r'},\omega) \approx
\sum_{\sigma=\uparrow \downarrow}
\sum_{q}^{Q}
\alpha_{q} \frac{\Phi_{q}({\bf r}) \Phi_{q}({\bf r'})}
                {\omega - \omega_{q \sigma}} \,.
\end{equation}
This truncation explicitly takes into account
numerous poles of the noninteracting response function.
In any adiabatic approximation to the XC kernel, the matrix elements
$M_{q \sigma \, q'  \sigma'}$ are real and frequency independent.
In this case the
excitation energies $\Omega$ are simply the
eigenvalues of the ($Q \times Q$) matrix
$ M_{q \sigma \, q'  \sigma'}(\Omega=0)
+ \delta_{q \sigma , q' \sigma'} \omega_{q \sigma} $.
For bound states of
finite systems we encounter well-separated poles
in the linear density response.  In our calculations, we include
many such poles, but only those of bound states, ignoring
continuum contributions.  The nature and size of the error this
introduces has been studied by van Gisbergen 
et al.\cite{GisbergenEtAl:98}, and does not affect the qualitative
conclusions found in this work.

A simple and extremely instructive case is when we expand
about a single KS-orbital energy
difference $\omega_{p \tau}$
\cite{PetersilkaGossmannGross:96,PetersilkaGross:96}.
The physical excitation energies $\Omega$ are then given by the
solution of
\begin{equation} \label{eqforexactomega}
\lambda(\Omega)
  =\frac{A(\omega_{p \tau})}{\Omega - \omega_{p \tau}} + B(\omega_{p \tau})
  + \ldots 
= 1 \,.
\end{equation}
For non-degenerate single-particle poles $\omega_{p \tau}$, the
coefficients in Eq.\ (\ref{eqforexactomega}) are given by
\begin{equation}
\label{Koeff_A}
A(\omega_{p \tau}) =  M_{p \tau \, p \tau}(\omega_{p \tau})
\end{equation}
and
\begin{equation}
\label{Koeff_B}
B(\omega_{p \tau}) =
\left.\frac{d M_{p \tau p \tau}}
           {d\omega}\right|_{\omega_{p \tau}}
+ \frac{1}{M_{p \tau p \tau}(\omega_{p \tau})}
\sum_{q' \sigma' \neq p \tau}
\frac{M_{p \tau \, q' \sigma'}(\omega_{p \tau})
      M_{q' \sigma' \, p \tau}(\omega_{p \tau}) }
     {\omega_{p \tau} - \omega_{q' \sigma'} + i\eta} \,.
\end{equation}
If the pole $\omega_{p \tau}$ is $\wp$-fold degenerate,
$
\omega_{p_1 \tau_1} = \omega_{p_2 \tau_2} = \ldots =
\omega_{p_{\wp} \tau_{\wp}} \equiv \omega_0 \,,
$
the lowest-order coefficient $A$ in Eq.~(\ref{eqforexactomega}) 
is determined by
a $\wp$-dimensional matrix equation
\begin{equation}
\label{excitC}
\sum_{k=1}^{\wp} M_{p_i \tau_i \, p_k \tau_k}(\omega_0)
\xi_{p_k \tau_k}^{(n)} = A_n(\omega_0)\xi^{(n)}_{p_i \tau_i}
\,, \quad i=1\ldots\wp \,,
\end{equation}
with $\wp$ different solutions $A_1 \ldots A_{\wp}$.
For excitation energies $\Omega$ close to
$\omega_{0}$, the
lowest-order term of the above Laurent expansion will dominate the
series.
In this single-pole approximation (SPA), the excitation energies
$\Omega$ satisfy
Eq.\ (\ref{eqforexactomega})
reduces to
\begin{equation}
\label{applic09}
\lambda_n(\Omega) \approx
\frac{A_n(\omega_{0})}{\Omega - \omega_{0}}
= 1 \,.
\end{equation}
The condition (\ref{lambda=1}) and its complex conjugate,
$\lambda^{*}(\Omega) = 1$, finally lead to a compact
expression for the excitation energies.

\begin{equation}
\label{excitF}
\Omega_n \approx \omega_0 + \Re A_n(\omega_0) \,.
\end{equation}

For closed-shell systems, every Kohn-Sham orbital eigenvalue is
degenerate with respect to spin, i.e.\ the spin multiplet
structure is absent in the bare Kohn-Sham eigenvalue spectrum.
Within the SPA, the dominant terms in the corrections to the Kohn-Sham
eigenvalues towards the true
multiplet energies naturally emerge from the solution of the
($2 \times 2$) eigenvalue problem
\begin{equation}
\label{applic10}
\sum_{\sigma'=\uparrow,\downarrow}
M_{p \sigma p \sigma'}(\omega_{0})
\xi_{p \sigma'}(\omega_{0})
=
A \xi_{p \sigma}(\omega_{0}) \,.
\end{equation}
Then, the resulting excitation energies are:
\begin{equation}
\label{applic25}
\Omega_{1,2} = \omega_{0}
+
\Re\left\{ M_{p \uparrow p \uparrow} \pm  M_{p \uparrow p \downarrow}
\right\}  \,.
\end{equation}
Using the explicit form of the matrix elements (\ref{excit9}) one
finds%
\footnote{Since we are dealing with spin saturated systems, we have
dropped the spin-index of $\Phi_{p \sigma}$.}
\bea
\label{applic40}
\Omega_1 &=& \omega_{0} + 2 \Re \int d^3r \int d^3 r' \,
\Phi^{*}_{p}({\bf r})
\left(
\frac{1}{|{\bf r}-{\bf r'}|}
+ f_{{\rm xc}}({\bf r},{\bf r'};\omega_0)
\right)
\Phi_{p}({\bf r'}) \\
\Omega_2 &=& \omega_{0} + 2 \Re \int d^3r \int d^3 r' \,
\Phi^{*}_{p}({\bf r})
G_{{\rm xc}}({\bf r},{\bf r'};\omega_0)
\Phi_{p}({\bf r'}) \,.
\label{applic50}
\eea
The kernel $G_{\rm xc}$
embraces the exchange and correlation effects in the
Kohn-Sham equation for the linear response of the
frequency-dependent magnetization density $m({\bf r},\omega)$
\cite{LiuVosko:89}.
For unpolarized systems, the weight of the
pole in the spin-summed susceptibility (both
for the Kohn-Sham and the physical systems)
at $\Omega_2$ is exactly zero, indicating
that these are the optically forbidden transitions to triplet
states.
The singlet excitation energies
are at $\Omega_1$. 
In this way, the SPA already gives rise to a spin-multiplet structure in the
excitation spectrum.  We use SPA to understand the results
of different approximations, since it simply relates the calculated
shifts from KS eigenvalues to matrix elements of the XC kernel.

At this point we stress that
the TDDFT formalism for the calculation of excitation 
energies involves three different types of approximations:
\begin{enumerate}
\item 
In the calculation of the Kohn-Sham orbitals $\varphi_{k}({\bf r})$
and their eigenvalues $\epsilon_k$,
one employs some approximation of the static XC potential $v_{\rm xc}$. 
\item 
Given the stationary Kohn-Sham orbitals and the ground state density,
the functional form of the XC kernel $f_{{\rm xc}\, \sigma \sigma'}$ 
needs to be approximated in order to calculate the matrix elements
defined in Eq.\ (\ref{excit9}).
\item 
Once the matrix elements are obtained,
the infinite-dimensional eigenvalue problem (\ref{matrixproblem}) (or,
equivalently, (\ref{eigenwertgl})) must be truncated in one way or
another.
\end{enumerate}

In the following, we are going to investigate the relative importance
of the approximations (1.) and (2.).
Furthermore, truncation effects will be
estimated by comparing the results obtained in SPA 
(\ref{applic40},\ref{applic50})
with the solution of the ``full'' problem (\ref{eigenwertgl}) which is based
on using up to 38 bound virtual orbitals.

\section{Results for the Helium Atom}
\label{section:Results_for_the_Helium_Atom}

In this section we report numerical results for excitation energies of
the He atom.
The stationary Kohn-Sham equations were solved numerically on a radial
grid (i.e. without basis set expansion)
using 
a large number of
semi-logarithmically distributed grid points 
\cite{ChernyshevaCherepkovRadojevic:76} up to a
maximum radius of several hundred atomic units in order 
to achieve high accuracy 
the Rydberg states ($n\geq10$) as well.

\subsection{Exact Kohn-Sham potential}
  
To eliminate the errors (1.) associated with the
approximation for the ground-state KS potential, we employ the {\em
  exact} XC potential of the He atom to generate the 
stationary Kohn-Sham orbitals $\varphi_{k}({\bf r})$ 
and their eigenvalues $\epsilon_k$. 
This isolates the effects which exclusively
arise due to the approximations
(2.) and (3.).
The potential data provided by Umrigar and
Gonze \cite{UmrigarGonze:94} were interpolated nonlinearly for $r\le10$
atomic units. Around $r=10$ atomic units, the XC potential is almost 
identical to $-1/r$. This behavior was used as an extrapolation of the
exact exchange-correlation potential to larger distances.
  
%
\begin{table}
\caption{\label{table:1}
  Singlet excitation energies of neutral helium,
  calculated from the exact XC potential by
  using approximate XC kernels (in atomic units)}
\begin{tabular}{ccccccccc}
\hline 
   &  &\multicolumn{2}{c}{\rm ALDA (xc)} 
      &\multicolumn{2}{c}{\rm TDOEP (x-only)} 
      &\multicolumn{2}{c}{\rm TDOEP (SIC)}\\ 
\cline{3-4} \cline{5-6} \cline{7-8}
  $ k \rightarrow j $  
&$\omega_{\rm jk}$
& SPA    & full\refnote aa   
& SPA    & full\refnote a  
& SPA    & full\refnote a  
& exact\refnote b\\ 
\hline
{$1s\rightarrow 2s$}& 0.7460 & 0.7718 & 0.7678 & 0.7687 & 0.7659 & 0.7674 &
0.7649 & 0.7578 \\
{$1s\rightarrow 3s$}& 0.8392 & 0.8458 & 0.8461 & 0.8448 & 0.8450 & 0.8445 &
0.8448 & 0.8425 \\
{$1s\rightarrow 4s$}& 0.8688 & 0.8714 & 0.8719 & 0.8710 & 0.8713 & 0.8709 &
0.8712 & 0.8701 \\
{$1s\rightarrow 5s$}& 0.8819 & 0.8832 & 0.8835 & 0.8830 & 0.8832 & 0.8829 &
0.8832 & 0.8825 \\
{$1s\rightarrow 6s$}& 0.8888 & 0.8895 & 0.8898 & 0.8894 & 0.8896 & 0.8894 &
0.8895 & 0.8892 \\
\hline
{$1s\rightarrow 2p$}& 0.7772 & 0.7764 & 0.7764 & 0.7850 & 0.7844 & 0.7836 &
0.7833 & 0.7799 \\
{$1s\rightarrow 3p$}& 0.8476 & 0.8483 & 0.8483 & 0.8500 & 0.8501 & 0.8497 &
0.8498 & 0.8486 \\
{$1s\rightarrow 4p$}& 0.8722 & 0.8726 & 0.8726 & 0.8732 & 0.8733 & 0.8731 &
0.8732 & 0.8727 \\
{$1s\rightarrow 5p$}& 0.8836 & 0.8838 & 0.8838 & 0.8841 & 0.8842 & 0.8841 &
0.8841 & 0.8838 \\
{$1s\rightarrow 6p$}& 0.8898 & 0.8899 & 0.8899 & 0.8901 & 0.8901 & 0.8900 &
0.8901 & 0.8899 \\
 \hline
{Mean abs. dev.\refnote{c}}
                    & 0.0022 & 0.0023 & 0.0021 & 0.0022 & 0.0020 & 0.0019 &
0.0017 \\
{Mean rel.\ dev.\refnote{d}}              
                    & 0.28\% & 0.30\% & 0.26\% & 0.28\% & 0.25\% & 0.24\% &
0.21\% \\
\hline
\end{tabular}
{
\tablenote{a} Using the lowest 34 unoccupied orbitals of s and p symmetry,
respectively.\\ 
\tablenote{b} Nonrelativistic variational calculation [38].\\
\tablenote{c} Mean value of the absolute deviations from the exact
              values.}
\end{table}

\begin{table}
\caption{\label{table:2}
  Triplet excitation energies of neutral helium,
  calculated from the exact XC potential by
  using approximate XC kernels (in atomic units)}
\begin{tabular}{ccccccccc}
\hline
   &  &\multicolumn{2}{c}{\rm ALDA (xc)} 
      &\multicolumn{2}{c}{\rm TDOEP (x-only)} 
      &\multicolumn{2}{c}{\rm TDOEP (SIC)}\\ 
\cline{3-4} \cline{5-6} \cline{7-8}
   $ k \rightarrow j $  
&$\omega_{jk}$
& SPA    & full\refnote a   
& SPA    & full\refnote a  
& SPA    & full\refnote a  
& exact\refnote b\\ 
\hline
{$1s\rightarrow 2s$}& 0.7460 & 0.7357 & 0.7351 & 0.7232 & 0.7207 & 0.7313 &
0.7300 & 0.7285\\
{$1s\rightarrow 3s$}& 0.8392 & 0.8366 & 0.8368 & 0.8337 & 0.8343 & 0.8353 &
0.8356 & 0.8350 \\
{$1s\rightarrow 4s$}& 0.8688 & 0.8678 & 0.8679 & 0.8667 & 0.8671 & 0.8673 &
0.8675 & 0.8672 \\
{$1s\rightarrow 5s$}& 0.8819 & 0.8814 & 0.8815 & 0.8808 & 0.8811 & 0.8811 &
0.8813 & 0.8811 \\
{$1s\rightarrow 6s$}& 0.8888 & 0.8885 & 0.8885 & 0.8882 & 0.8883 & 0.8884 &
0.8884 & 0.8883 \\
\hline
{$1s\rightarrow 2p$}& 0.7772 & 0.7702 & 0.7698 & 0.7693 & 0.7688 & 0.7774 &
0.7774 & 0.7706 \\
{$1s\rightarrow 3p$}& 0.8476 & 0.8456 & 0.8457 & 0.8453 & 0.8453 & 0.8471 &
0.8471 & 0.8456 \\
{$1s\rightarrow 4p$}& 0.8722 & 0.8714 & 0.8715 & 0.8712 & 0.8713 & 0.8720 &
0.8720 & 0.8714 \\
{$1s\rightarrow 5p$}& 0.8836 & 0.8832 & 0.8832 & 0.8831 & 0.8831 & 0.8834 &
0.8835 & 0.8832 \\
{$1s\rightarrow 6p$}& 0.8898 & 0.8895 & 0.8895 & 0.8895 & 0.8895 & 0.8897 &
0.8897 & 0.8895 \\
 \hline
{Mean abs. dev.\refnote{c}}  
                    & 0.0035 & 0.0010 & 0.0011 & 0.0009 & 0.0011 & 0.0013 &
0.0012 \\
{Mean rel.\ dev.}
                    & 0.45\% & 0.14\% & 0.14\% & 0.12\% & 0.15\% & 0.16\% &
0.15\% \\
\hline
\end{tabular}
{
\tablenote{a} Using the lowest 34 unoccupied orbitals of s and p symmetry,
respectively.\\ 
\tablenote{b} Nonrelativistic variational calculation [38].\\
\tablenote{c} Mean value of the absolute deviations from the exact
values.}
\end{table}

Tables \ref{table:1} and \ref{table:2} show the excitation energies of
neutral helium calculated with the exact exchange-correlation potential.
The results are compared with a highly accurate
nonrelativistic variational calculation \cite{KonoHattori:84}
of the eigenstates of Helium.
It is a remarkable fact that
the Kohn-Sham excitation energies $\omega_{jk}=\epsilon_j-\epsilon_k$ 
are already very close to the
exact spectrum, and, at the same time, 
are always in between the singlet and the triplet 
energies\cite{SUG98,BPG00}.

Based on these eigenvalue differences, we have calculated the shifts
towards the true excitation energies using
several approximations for the exchange-correlation kernels $f_{\rm
  xc}$:
\begin{itemize}
\item
The adiabatic local density approximation (ALDA), 
with the inclusion of correlation contributions in the 
parametrization of Vosko, Wilk and Nusair
\cite{VoskoWilkNusair:80}.
\item
The approximate X-only time-dependent OEP (TDOEP) kernel of Eq.\
(\ref{fxcoep}), which is based on the time-dependent Fock expression,
and
\item 
The approximate TDOEP-SIC kernel from Eq.\
(\ref{fxc_oepsic}) with the parametrization  
of Ref.\ \cite{VoskoWilkNusair:80} for the correlation contributions.
\end{itemize}

The columns denoted by ``full'' show the corresponding excitation
energies $\Omega_i$ which are obtained as eigenvalues 
obtained from the (truncated) matrix equation (\ref{eigenwertgl}).
To  investigate the effects
of the truncation of the matrix equation (\ref{eigenwertgl})
we compare the difference between the single-pole approximation (SPA)
and the fully coupled results.
The matrix equation (\ref{eigenwertgl}) was solved using
$N=34$ unoccupied
Kohn-Sham orbitals of $s$ or $p$ symmetry.
For each symmetry class
the resulting dimension of the (fully coupled but
truncated)  matrix in Eq.\ (\ref{eigenwertgl}) is $(4N \times 4N)$
(due to the spin-degeneracy of the KS orbitals of Helium  
and the fact that the frequency-dependent Kohn-Sham response
function is symmetric in the complex plane with respect to the
imaginary axis). Thereby, convergence of the results
to within $10^{-6}$ atomic units was reached
within the space of bound states. 
When comparing the results from the SPA with the results from
the fully coupled matrix,
we observe only a small change
in the resulting excitation energies
(from a few hundredth of a percent to at most one half percent),
independent of the functional form of the exchange-correlation kernel.
Thus we conclude that
in helium the single-pole approximation gives the
dominant correction to the Kohn-Sham excitation spectrum.
Hence, starting from the Kohn-Sham eigenvalue differences as zeroth
order approximation to the excitation energies, the SPA can be used
for the assignment of the excitation energies which 
are obtained as eigenvalues from Eq.\ (\ref{eigenwertgl}).
Recent studies using basis set expansions \cite{GisbergenEtAl:98}
indicate that further improvement of the fully coupled results
can be expected from the inclusion of continuum states. 
The general trends of the results however, are not affected.

\begin{figure}[ht!]
\unitlength1.0cm
\begin{picture}(15.4,14)
\put(-5.25,-1){\makebox(12,6){
\includegraphics{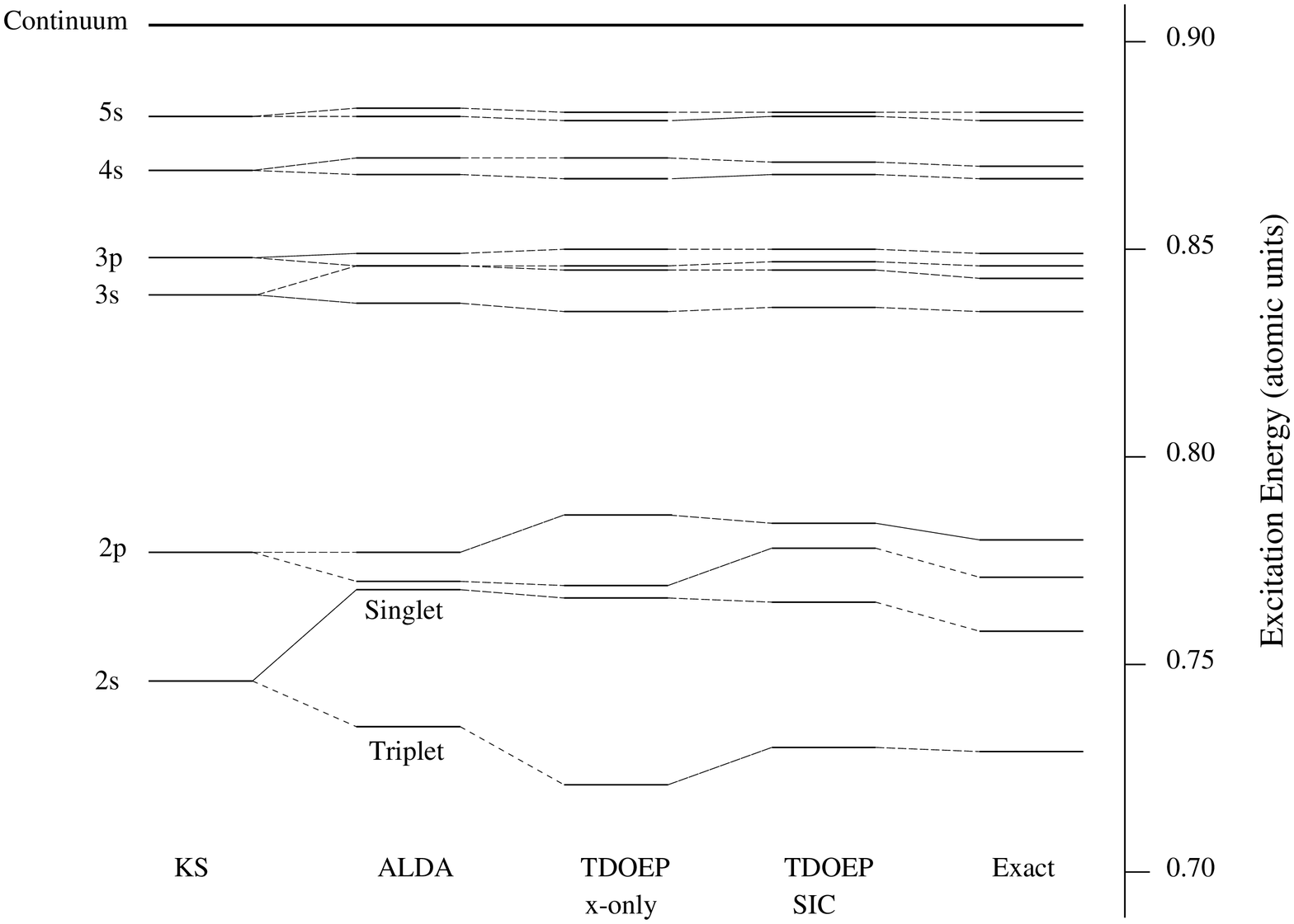}
}}
\end{picture}
\caption{\small\label{fig:2} Typical excitation energies of He, including
  the orbital eigenvalues of the exact Kohn-Sham potential (KS) and
  the corrections from time-dependent density functional theory
  calculated within the adiabatic local density approximation (ALDA),
  with orbital dependent functionals in the X-only limit 
  (TDOEP X-only) and
  the self-interaction corrected version of the ALDA (TDOEP-SIC).
}
\end{figure}

In figure \ref{fig:2} we have plotted some typical excitation
energies taken from the column headed ``full'' of
table \ref{table:1} and \ref{table:2},
%
We can understand the trends in this figure by analyzing the
results in terms of the single-pole approximation.
For the single-particle excitations in helium, the single-pole
approximation leads to two-dimensional
matrix equations for the excitation energies (c.f. Eqs.\
(\ref{applic10}) - (\ref{applic40})). 
In the following, the notation  
\begin{equation} 
\langle \hat{\cal O} \rangle := \int d^3r \int d^3 r' \,
\Phi^{*}_{p}({\bf r})
\hat{\cal O}({\bf r},{\bf r'})
\Phi_{p}({\bf r'}) 
\end{equation}
will be used
for the matrix elements of the two particle operators $\hat{\cal O}$
involved in the calculation.
Then, in the SPA,
\be
\Omega_p^{\rm singlet} 
= \omega_p + 2\langle W \rangle + 2 \langle f_{\rm xc} \rangle,~
\Omega_p^{\rm triplet} 
= \omega_p + 2 \langle  G_{\rm xc} \rangle,~
\Delta \Omega_p =
2 (\langle W \rangle + \langle f_{\rm c} \rangle - \langle  G_{\rm c}
\rangle) \label{levelSPA},
\ee
where $\Delta\Omega_p$ is the singlet-triplet splitting.
Within the various approximations to the kernel, these levels become
\bea
\Omega_p^{\rm sing} 
&=&\omega_p + \langle W \rangle,~~~~~~~~~~~~~~~~~~~~~
\Omega_p^{\rm triplet} 
= \omega_p - \langle W \rangle,~~~~~~~
~~~~~~({\rm X-only})
\\
&=&\omega_p + 2\langle W \rangle + 2 \langle f_{\rm xc}^{\rm ALDA} \rangle,~
~~~~~~~~~~~
= \omega_p + 2 \langle  G_{\rm xc}^{\rm ALDA} \rangle,~
~~~~~~~~~~{\rm ALDA}
\\
&=& \omega_p + \langle W \rangle + 2 \langle f_{\rm c}^{\rm ALDA} \rangle
-\langle f_{\rm c}^{\rm orb} \rangle,~
= \omega_p - \langle W \rangle
+ 2 \langle  G_{\rm c}^{\rm ALDA} \rangle
-\langle f_{\rm c}^{\rm orb} \rangle.~
{\rm SIC}
\label{levels}
\eea

We begin our analysis with the splitting.
In the simplest case, the TDOEP X-only kernel,
we see that the singlet transitions are always overestimated, while the
triplets are always underestimated.   Since our TDOEP treatment is
exact for exchange in this case, this underscores the importance
of correlation.   In particular, since $\langle f_{\rm x} \rangle
=\langle G_{\rm x} \rangle = - \langle W \rangle/2$, the splitting is
just $2\langle W \rangle$.
This matrix element is always positive, correctly
putting the singlet above the
triplet,  but the splitting
is typically far too big.
We demonstrate the effect of this in table 3, in which we compare
splittings with and without correlation.
\begin{table}
\caption{\label{table:3}
  Singlet-triplet separations in helium
  obtained from Eq.\ (\protect\ref{eigenwertgl}),
  using the lowest 34 
  unoccupied orbitals of s and p symmetry
  of the exact XC potential and
  employing various approximate XC kernels. 
  All values are in mHartrees.}
\begin{tabular}{crrrrr}
\hline
 &\multicolumn{2}{c}{\rm ALDA} 
 &\multicolumn{2}{c}{\rm TDOEP} \\ \cline{2-3} \cline{4-5}
State & X-only & xc\refnote a
      & X-only          & SIC    
      & exact\refnote b\\ 
\hline
2$S$& 42.2 & 32.7 & 45.2  & 34.9 & 29.3 \\
3$S$& 11.1 &  9.4 & 10.8  & 9.2  &  7.4 \\
4$S$&  4.7 &  4.0 &  4.3  & 3.7  &  2.9 \\
5$S$&  2.4 &  2.1 &  2.2  & 1.9  &  1.4 \\
6$S$&  1.4 &  1.3 &  1.2  & 1.1  &  0.8 \\
\hline
2$P$& 16.7 & 6.6 & 15.6 & 5.8 &  9.3 \\
3$P$&  4.5 & 2.6 & 4.7  & 2.7 &  2.9 \\
4$P$&  1.8 & 1.1 & 2.0  & 1.2 &  1.3 \\
5$P$&  0.9 & 0.6 & 1.0  & 0.7 &  0.6 \\
6$P$&  0.5 & 0.3 & 0.6  & 0.4 &  0.4 \\
\hline
{dev.\refnote c}& 3.0 & 1.1 & 3.1 & 1.3 \\
{rel.\ dev.}    &55\% &27\% &56\% &21\% \\
\hline\\
\end{tabular} 
\newline \tablenote{a} Including correlation contributions in the form of
Vosko, Wilk and Nusair {\protect\cite{VoskoWilkNusair:80}}.\\
\tablenote{b} Taken from Ref.\ {\protect\cite{KonoHattori:84}}\\
\tablenote{c} Mean absolute deviation from the exact values
\end{table}
To see why inclusion of correlation always reduces the splitting, we note
the sign and magnitude of matrix elements, within ALDA.  Even though
both $\langle f_{\rm xc}^{\rm ALDA} \rangle$ and
$\langle G_{\rm xc}^{\rm ALDA} \rangle$ are negative, because they are
dominated by their exchange contributions, we find
\be
\langle f_{\rm c}^{\rm ALDA} \rangle 
<
-\langle  G_{\rm c}^{\rm ALDA} \rangle 
< 0
\ee
because in Eq. (\ref{fxcGxc})
antiparallel correlation dominates over parallel correlation.
Thus the ALDA correlation contribution to the splitting
is always negative in SPA.
Note that the SIC treatment of the splitting is only marginally better 
than in ALDA because, within SPA, 
the SIC splitting is identical to that of ALDA.

To analyze the separate levels, we need the magnitude of the SIC corrections:
\begin{equation} \label{both_are_close}
\langle f_{\rm c}^{\rm ALDA} \rangle 
<
\langle f_{\rm c}^{\rm orb} \rangle 
:=
\langle \frac{\delta v_{\rm c}^{\rm LDA}[n_{k\sigma},0]}
             {\delta n_{k\sigma}} 
\rangle
< 0 \,,
\end{equation}
but the numerical values of both matrix elements differ by less than 8\%. 
Moreover, 
\begin{equation} \label{Gc>fc}
\langle  G_{\rm c}^{\rm ALDA} \rangle 
>
| \langle f_{\rm c}^{\rm orb} \rangle |
> 0 \,.
\end{equation}
Looking at the {\em singlet} excitation energies of table \ref{table:1}
we see that in ALDA, the s-levels are too high (up to 10 mH), whereas
the p-levels are too low (by up to 0.4 mH). In X-only
TDOEP, the s-levels drop (by up to 3 mH), approaching the exact values, 
but the rise of the p-states (by up to 8mH) 
is too high. Incorporating explicit correlation terms by using the
TDOEP-SIC kernel, the singlet lines correctly
drop further (in comparison to the X-only results by up to 1 mH) since 
$2\langle f_{\rm c}^{\rm ALDA} \rangle-
\langle f_{\rm c}^{\rm orb} \rangle \approx  
\langle f_{\rm c}^{\rm ALDA} \rangle$ in Eq.\ (\ref{levels}) is always
a negative contribution. But still, the p-states are too high.
Regarding the {\em triplet} excitation energies of table \ref{table:2},
the ALDA s-states are 
too high by at most 6mH, but the p-states are almost identical
to the exact values. In X-only TDOEP, the
triplet states experience a strong downshift from the Kohn-Sham
excitation energies up to 25 mH, originating from the term
$-\langle W \rangle$ (see Eq.\ (\ref{levels})).
In TDOEP-SIC, this downshift is partly screened by the positive correlation 
contributions $2\langle G_{\rm c}^{\rm ALDA} \rangle - 
\langle f_{\rm c}^{\rm orb}
\rangle$, as can be seen from Eqs.\ (\ref{levels}),
(\ref{both_are_close}) and (\ref{Gc>fc}). This leads to an excellent
agreement with the exact values for the s-states. However, 
these correlation terms are too large for the p-states.
Since 
$\langle G_{\rm c}^{\rm ALDA} \rangle > 
\langle f_{\rm c}^{\rm ALDA} \rangle$, the rise
of the triplet is always bigger than the dropping of the singlet.

\subsection{Approximate Kohn-Sham potentials}

Next we explore the effect of approximate 
exchange-correlation potentials $v_{\rm xc}$ 
on the calculated excitation spectrum of the He atom.
We do not even report results
within LDA and generalized gradient
approximations (GGA)\cite{EngelChevaryMacdonaldVosko:92},
since these potentials only support a few virtual states, so that
many of the transitions reported here do not even exist in such
calculations.  (This problem is worst in small atoms, is
less pronounced in molecules, and irrelevant in solids).

To produce a  correct Rydberg series,  the XC potential must decay
as $-1/r$, an exact exchange effect.
Hence we examine the
 OEP X-only potential (which, {\em for two-electron systems} 
is identical to the Hartree-Fock potential) and the
OEP-SIC \cite{ChenEtAl:96} potential.
Both potentials show the
correct behavior for large distances from the nucleus, and
support of all the Rydberg states is guaranteed.
\begin{table}
\caption{\label{table:4}
Singlet excitation energies of neutral helium,
  calculated from the X-only potential and by
  using approximate XC kernels (in atomic units).}
\begin{tabular}{ccccccc}
\hline
&   &\multicolumn{2}{c}{ ALDA (xc)} &
    \multicolumn{2}{c}{\rm TDOEP (X-only)} \\
\cline{3-4} \cline{5-6}
Transition &  
$\omega_{jk}$ &
SPA    & full\refnote a  &
SPA    & full\refnote a  &
exact\refnote b\\ 
\hline
{$1s\rightarrow 2s$}& 0.7596 & 0.7852 & 0.7812 & 0.7822 & 0.7794 & 0.7578 \\
{$1s\rightarrow 3s$}& 0.8533 & 0.8598 & 0.8601 & 0.8588 & 0.8591 & 0.8425 \\
{$1s\rightarrow 4s$}& 0.8830 & 0.8856 & 0.8860 & 0.8851 & 0.8855 & 0.8701 \\
{$1s\rightarrow 5s$}& 0.8961 & 0.8973 & 0.8977 & 0.8971 & 0.8974 & 0.8825 \\
{$1s\rightarrow 6s$}& 0.9030 & 0.9037 & 0.9040 & 0.9036 & 0.9038 & 0.8892 \\
\hline
{$1s\rightarrow 2p$}& 0.7905 & 0.7900 & 0.7900 & 0.7986 & 0.7981 & 0.7799 \\
{$1s\rightarrow 3p$}& 0.8616 & 0.8623 & 0.8623 & 0.8640 & 0.8641 & 0.8486 \\
{$1s\rightarrow 4p$}& 0.8864 & 0.8867 & 0.8867 & 0.8874 & 0.8875 & 0.8727 \\
{$1s\rightarrow 5p$}& 0.8978 & 0.8980 & 0.8980 & 0.8983 & 0.8984 & 0.8838 \\
{$1s\rightarrow 6p$}& 0.9040 & 0.9041 & 0.9041 & 0.9043 & 0.9043 & 0.8899 \\
\hline
{Mean abs. dev.\refnote{c}}  & 0.0118 & 0.0156 & 0.0153 & 0.0162 & 0.0161 \\
{Mean percentage error}      & 1.37\% & 1.85\% & 1.81\% & 1.93\% & 1.90\% \\
\hline
\end{tabular}
\newline%
\tablenote{a} Using the lowest 34 unoccupied orbitals of s and p symmetry,
respectively.\\ 
\tablenote{b} Nonrelativistic variational calculation [38].\\
\tablenote{c} Mean value of the absolute deviations from the exact
values.
\end{table}

\begin{table}
\caption{\label{table:5}
Triplet excitation energies of neutral helium,
  calculated from the X-only potential and by
  using approximate XC kernels (in atomic units).}
\begin{tabular}{ccccccc}
\hline
&   &\multicolumn{2}{c}{ ALDA (xc)} &
    \multicolumn{2}{c}{\rm TDOEP (X-only)} \\
\cline{3-4} \cline{5-6}
Transition &  
$\omega_{jk}$ &
SPA    & full\refnote a  &
SPA    & full\refnote a  &
exact\refnote b\\ 
\hline
{$1s\rightarrow 2s$}& 0.7596 & 0.7493 & 0.7488 & 0.7370 & 0.7345 & 0.7285 \\
{$1s\rightarrow 3s$}& 0.8533 & 0.8507 & 0.8508 & 0.8478 & 0.8484 & 0.8350 \\
{$1s\rightarrow 4s$}& 0.8830 & 0.8820 & 0.8821 & 0.8809 & 0.8812 & 0.8672 \\
{$1s\rightarrow 5s$}& 0.8961 & 0.8956 & 0.8957 & 0.8950 & 0.8953 & 0.8811 \\
{$1s\rightarrow 6s$}& 0.9030 & 0.9027 & 0.9028 & 0.9024 & 0.9026 & 0.8883 \\
\hline
{$1s\rightarrow 2p$}& 0.7905 & 0.7833 & 0.7830 & 0.7824 & 0.7819 & 0.7706 \\
{$1s\rightarrow 3p$}& 0.8616 & 0.8595 & 0.8596 & 0.8591 & 0.8592 & 0.8456 \\
{$1s\rightarrow 4p$}& 0.8864 & 0.8855 & 0.8856 & 0.8853 & 0.8854 & 0.8714 \\
{$1s\rightarrow 5p$}& 0.8978 & 0.8973 & 0.8974 & 0.8972 & 0.8973 & 0.8832 \\
{$1s\rightarrow 6p$}& 0.9040 & 0.9037 & 0.9037 & 0.9037 & 0.9037 & 0.8895 \\
\hline
{Mean abs. dev.\refnote{c}}  & 0.0175 & 0.0149 & 0.0149 & 0.0130 & 0.0129 \\
{Mean percentage error}      & 2.11\% & 1.78\% & 1.78\% & 1.53\% & 1.51\% \\
\hline
\end{tabular}
{\newline%
\tablenote{a} Using the lowest 34 unoccupied orbitals of s and p symmetry,
respectively.\\ 
\tablenote{b} Nonrelativistic variational calculation [38].\\
\tablenote{c} Mean value of the absolute deviations from the exact
values.}
\end{table}

\begin{table}
\caption{\label{table:6} Singlet excitation energies of neutral helium,
  calculated from the SIC-LDA potential and by
  using approximate XC kernels (in atomic units).}
{
\begin{tabular}{ccccccc}
\hline
&   &\multicolumn{2}{c}{ ALDA (xc)} &
    \multicolumn{2}{c}{\rm TDOEP (SIC)} \\
\cline{3-4} \cline{5-6}
Transition &  
$\omega_{jk}$ &
SPA    & full\refnote a  &
SPA    & full\refnote a  &
exact\refnote b\\ 
\hline
{$1s\rightarrow 2s$}& 0.7838 & 0.8111 & 0.8070 & 0.8065 & 0.8039 & 0.7578 \\
{$1s\rightarrow 3s$}& 0.8825 & 0.8891 & 0.8895 & 0.8878 & 0.8881 & 0.8425 \\
{$1s\rightarrow 4s$}& 0.9130 & 0.9156 & 0.9161 & 0.9150 & 0.9154 & 0.8701 \\
{$1s\rightarrow 5s$}& 0.9263 & 0.9276 & 0.9280 & 0.9273 & 0.9276 & 0.8825 \\
{$1s\rightarrow 6s$}& 0.9333 & 0.9340 & 0.9343 & 0.9339 & 0.9341 & 0.8892 \\
\hline
{$1s\rightarrow 2p$}& 0.8144 & 0.8145 & 0.8144 & 0.8222 & 0.8217 & 0.7799 \\
{$1s\rightarrow 3p$}& 0.8906 & 0.8915 & 0.8915 & 0.8929 & 0.8930 & 0.8486 \\
{$1s\rightarrow 4p$}& 0.9163 & 0.9167 & 0.9167 & 0.9172 & 0.9173 & 0.8727 \\
{$1s\rightarrow 5p$}& 0.9280 & 0.9282 & 0.9282 & 0.9285 & 0.9285 & 0.8838 \\
{$1s\rightarrow 6p$}& 0.9343 & 0.9344 & 0.9344 & 0.9346 & 0.9346 & 0.8899 \\
\hline
{Mean abs. dev.\refnote{c}}  & 0.0406 & 0.0446 & 0.0443 & 0.0449 & 0.0447 \\
{Mean percentage error}      & 4.74\% & 5.25\% & 5.21\% & 5.29\% & 5.26\% \\
\hline
\end{tabular}
}
{\newline%
\tablenote{a} Using the lowest 34 unoccupied orbitals of s and p symmetry,
respectively.\\ 
\tablenote{b} Nonrelativistic variational calculation [38].\\
\tablenote{c} Mean value of the absolute deviations from the exact
values.}
\end{table}

\begin{table}
\caption{\label{table:7} Triplet excitation energies of neutral helium,
  calculated from the SIC-LDA potential and by
  using approximate XC kernels (in atomic units).}
{
\begin{tabular}{ccccccc}
\hline
&   &\multicolumn{2}{c}{ ALDA (xc)} &
    \multicolumn{2}{c}{\rm TDOEP (SIC)} \\
\cline{3-4} \cline{5-6}
Transition &  
$\omega_{jk}$ &
SPA    & full\refnote a  &
SPA    & full\refnote a  &
exact\refnote b\\ 
\hline
{$1s\rightarrow 2s$}& 0.7838 & 0.7727 & 0.7722 & 0.7681 & 0.7668 & 0.7285 \\
{$1s\rightarrow 3s$}& 0.8825 & 0.8799 & 0.8800 & 0.8786 & 0.8789 & 0.8350 \\
{$1s\rightarrow 4s$}& 0.9130 & 0.9120 & 0.9121 & 0.9115 & 0.9117 & 0.8672 \\
{$1s\rightarrow 5s$}& 0.9263 & 0.9258 & 0.9259 & 0.9256 & 0.9257 & 0.8811 \\
{$1s\rightarrow 6s$}& 0.9333 & 0.9331 & 0.9331 & 0.9329 & 0.9330 & 0.8883 \\
\hline
{$1s\rightarrow 2p$}& 0.8144 & 0.8062 & 0.8058 & 0.8140 & 0.8139 & 0.7706 \\
{$1s\rightarrow 3p$}& 0.8906 & 0.8885 & 0.8885 & 0.8899 & 0.8899 & 0.8456 \\
{$1s\rightarrow 4p$}& 0.9163 & 0.9154 & 0.9154 & 0.9159 & 0.9159 & 0.8714 \\
{$1s\rightarrow 5p$}& 0.9280 & 0.9275 & 0.9276 & 0.9278 & 0.9278 & 0.8832 \\
{$1s\rightarrow 6p$}& 0.9343 & 0.9340 & 0.9340 & 0.9342 & 0.9342 & 0.8895 \\
\hline
{Mean abs. dev.\refnote{c}}  & 0.0462 & 0.0435 & 0.0434 & 0.0438 & 0.0437 \\
{Mean percentage error}      & 5.50\% & 5.15\% & 5.14\% & 5.19\% & 5.18\% \\
\hline
\end{tabular}
} 
{\newline
\tablenote{a} Using the lowest 34 unoccupied orbitals of s and p symmetry,
respectively.\\ 
\tablenote{b} Nonrelativistic variational calculation [38].\\
\tablenote{c} Mean value of the absolute deviations from the exact
values.}
\end{table}

Tables \ref{table:4} and \ref{table:5} show the approximate Kohn-Sham
excitation energies and the corresponding corrected excitation
energies calculated from the approximate Kohn-Sham eigenvalues and
orbitals of the X-only potential; tables \ref{table:6} and \ref{table:7}
are their analogs from the OEP-SIC calculation. 
The Kohn-Sham orbital energy differences are
almost uniformly shifted to larger values
compared to the orbital energy differences of the exact Kohn-Sham
potential. The shift ranges from 13.6 mH for the lowest excitation
energy to 14.2 mH for excitation energies $\Omega_n$ with $n\geq4$
for the X-only potential.
The latter shift is exactly the difference between the exact 1s eigenvalue
($\epsilon_{1s}^{\rm exact}=-0.90372$ a.u.) and the more strongly bound 
1s eigenvalue of the X-only potential 
($\epsilon_{1s}^{\rm X-only}=-0.91796$ a.u.).
Similarly, the Kohn-Sham eigenvalue differences calculated in
OEP-SIC are shifted by up to 44.5 mH, which again is equal to
the difference between the 1s
eigenvalues of the exact Kohn-Sham potential and the KS potential in 
OEP-SIC.  In OEP-SIC, the correlation potential is attractive at all
points in space. Hence, including SIC-correlation contributions
into the OEP worsens the occupied orbital eigenvalue.
To summarize, the inclusion of correlation contributions to the
ground state potential mostly affects only the occupied state;
the virtual states are almost exact, i.e., they are
almost independent of the choice of the correlation potential.
The He Kohn-Sham orbitals exhibit a Rydberg-like behavior already for 
relatively low quantum numbers $n$ \cite{PetersilkaGossmannGross:98}:
already the lower virtual states
are mostly determined by the large-$r$ behavior of the Kohn-Sham
potential, which is governed by the exchange contribution.

As a consequence, the
corrections to the Kohn-Sham orbital energy differences,
calculated on the approximate orbitals,
are very close 
to the corrections calculated from the exact Kohn-Sham orbitals. 
This is most apparent from the singlet-triplet splittings
given in tables \ref{table:3} and \ref{table:8}: the splittings depend
more strongly on the choice of the XC kernel than on the choice of the
potential. 
\begin{table}[t]
\caption{\small\label{table:8}
  Singlet-triplet separations in neutral helium
  calculated from the X-only potential 
  and the SIC-potential and by 
  using various approximate XC kernels. 
  Calculated from Eq.\ (\protect\ref{eigenwertgl}), 
  using the lowest 34 
  unoccupied orbitals of s and p symmetry.
  All values are in mHartrees.}
\begin{tabular}{crrrrr}
\hline
 &\multicolumn{2}{c} {x-only} &
  \multicolumn{2}{c} {SIC}    &\\ \cline{2-3} \cline{4-5}
State & $f_{\rm xc}^{\rm ALDA}$ & $f_{\rm x}^{\rm TDOEP}$
      & $f_{\rm xc}^{\rm ALDA}$ & $f_{\rm xc}^{\rm TDOEP-SIC}$
      & exact\refnote b\\ 
\hline
2$S$ & 32.5 & 44.9 & 34.9 & 37.1 & 29.3 \\
3$S$ &  9.3 & 10.7 &  9.5 &  9.3 & 7.4 \\
4$S$ &  4.0 &  4.2 &  4.0 &  3.7 & 2.9 \\
5$S$ &  2.1 &  2.1 &  2.1 &  1.9 & 1.4 \\
6$S$ &  1.2 &  1.2 &  1.2 &  1.1 & 0.8 \\
\hline
2$P$ &  7.1 & 16.2 & 8.6 & 7.8 & 9.3 \\
3$P$ &  2.7 &  4.9 & 3.0 & 3.1 & 2.9 \\
4$P$ &  1.2 &  2.1 & 1.3 & 1.4 & 1.3 \\
5$P$ &  0.6 &  1.1 & 0.6 & 0.7 & 0.6 \\
6$P$ &  0.3 &  0.6 & 0.4 & 0.4 & 0.4 \\
\hline
{dev.\refnote c}& 1.46 & 4.26 & 1.98 & 2.26 \\
{rel.\ dev.}    & 35\% & 49\% & 37\% & 31\% \\
\hline\\
\end{tabular} 
\newline \tablenote{a} Including correlation contributions in the form of
Vosko, Wilk and Nusair {\protect\cite{VoskoWilkNusair:80}}.\\
\tablenote{b} Taken from Ref.\ {\protect\cite{KonoHattori:84}}\\
\tablenote{c} Mean absolute deviation from the exact values
\end{table}
However, for the excitation energies,
the differences among the various approximations of
the exchange correlation kernel are {\em smaller} than the differences 
in the Kohn-Sham excitation energies coming from different potentials. 
This reflects the fact that the resulting orbitals are rather insensitive to
different approximations of the potential.
Hence, the corrections themselves, calculated with approximate XC {\em
  kernels} will not cancel the shortcomings of an
approximate exchange 
{\em potential}. Tables \ref{table:4} - \ref{table:7} show that the
corrections go in the right direction only for the singlet states,
which are always lower than the corresponding Kohn-Sham orbital energy
differences. 
In other approximations, like the LDA and in the popular GGAs for
instance, this will be even more severe:
There the highest occupied orbital eigenvalue is in error 
by about a factor of two, due to spurious self-interaction.
There may be error cancellations for the lower
Kohn-Sham eigenvalue differences,
but in general one should not expect to get a reliable (Kohn-Sham)
spectrum in LDA and GGAs, because the respective potentials have 
the wrong behavior for large $r$.
In addition, this causes the number of (unoccupied) bound KS states
to be {\em finite}.

In total, the inaccuracies introduced by {\em approximate} ground
state Kohn-Sham potentials are substantial, but mostly reside in the
occupied eigenvalue for He.
It is very unlikely that these defects will be cured by better
approximations
of $f_{\rm xc}$ alone, since the terms containing $f_{\rm xc}$ only
give corrections to the underlying Kohn-Sham eigenvalue spectrum.
Hence, the quantitative calculation of excitation
energies heavily depends on the accuracy of the ground-state potential
employed.

\section{Results for the Beryllium Atom}
\label{section:Results_for_the_Beryllium_Atom}

\subsection{Exact Kohn-Sham potential}

The beryllium atom serves as a further 
standard example for first principles treatments:
besides numerous quantum chemical studies (e.g.\ 
\cite{Zuelicke:85,GrahamEtAl:86}),
a highly accurate ground-state exchange-correlation potential, obtained
from quantum Monte-Carlo methods \cite{UmrigarGonze:93}, is
available for this system.
With this potential,
we calculated accurate Kohn-Sham
orbitals and orbital energies of the beryllium atom.
In each symmetry class (s, p, and d), up to 38 virtual states were
calculated on a radial grid similar to the one used in section 
\ref{section:Results_for_the_Helium_Atom}.

In tables \ref{table:9} and \ref{table:10} we report the excitation
energies for the 11 lowest excitations of singlet and triplet
symmetry.
As in helium, the orbital energies of the accurate potential lie
always in between the experimental singlet and triplet energies.
However, the experimentally measured singlet-triplet separations
in beryllium are much larger than in the helium atom (cf.\ 
the last columns given in tables \ref{table:3} and \ref{table:11}).
Accordingly, to achieve agreement with the experimental data,
appreciable shifts of the Kohn-Sham eigenvalue differences 
are needed.

%
\begin{table}[t]
\caption{\label{table:9}
  Singlet excitation energies for the Be atom,
  calculated from the exact XC potential by
  using approximate XC kernels (in atomic units)}
\begin{tabular}{ccccccccc}
\hline 
   &  &\multicolumn{2}{c}{\rm ALDA (xc)} 
      &\multicolumn{2}{c}{\rm TDOEP (x-only)} 
      &\multicolumn{2}{c}{\rm TDOEP (SIC)}\\ 
\cline{3-4} \cline{5-6} \cline{7-8}
  $ k \rightarrow j $  
&$\omega_{jk}$
& SPA    & full\refnote a   
& SPA    & full\refnote a  
& SPA    & full\refnote a  
& Expt. \refnote b\\ 
\hline
{$2s\rightarrow 2p$}& 0.1327 & 0.2078 & 0.1889 & 0.2040 & 0.1873 & 0.2013 &
0.1855 & 0.1939 \\
{$2s\rightarrow 3s$}& 0.2444 & 0.2526 & 0.2515 & 0.2574 & 0.2553 & 0.2566 &
0.2547 & 0.2491 \\
{$2s\rightarrow 3p$}& 0.2694 & 0.2690 & 0.2714 & 0.2748 & 0.2758 & 0.2739 &
0.2750 & 0.2742 \\
{$2s\rightarrow 3d$}& 0.2833 & 0.2783 & 0.2779 & 0.2851 & 0.2851 & 0.2843 &
0.2842 & 0.2936 \\
{$2s\rightarrow 4s$}& 0.2959 & 0.2983 & 0.2984 & 0.2994 & 0.2995 & 0.2993 &
0.2994 & 0.2973 \\
{$2s\rightarrow 4p$}& 0.3046 & 0.3045 & 0.3049 & 0.3063 & 0.3067 & 0.3061 &
0.3065 & 0.3063 \\
{$2s\rightarrow 4d$}& 0.3098 & 0.3084 & 0.3084 & 0.3106 & 0.3106 & 0.3104 &
0.3103 & 0.3134 \\
{$2s\rightarrow 5s$}& 0.3153 & 0.3163 & 0.3164 & 0.3168 & 0.3170 & 0.3167 &
0.3169 & 0.3159 \\
{$2s\rightarrow 5p$}& 0.3193 & 0.3192 & 0.3194 & 0.3201 & 0.3203 & 0.3200 &
0.3202 & 0.3195 \\
{$2s\rightarrow 6s$}& 0.3247 & 0.3252 & 0.3253 & 0.3254 & 0.3256 & 0.3254 &
0.3256 & 0.325  \\
{$2s\rightarrow 6p$}& 0.3269 & 0.3268 & 0.3269 & 0.3273 & 0.3274 & 0.3272 &
0.3273 & 0.327  \\
 \hline
{Mean abs. dev.\refnote{c}}
             & 0.0081 & 0.0043 & 0.0031 & 0.0031 & 0.0028 & 0.0029 & 0.0029
\\
{Mean rel.\ dev.}              
             & 3.75\% & 1.69\% & 1.15\% & 1.27\% & 1.09\% & 1.13\% & 1.15\%
\\
\hline
{abs. dev.\refnote{d}}
             & 0.0028 & 0.0033 & 0.0029 & 0.0025 & 0.0025 & 0.0024 & 0.0024
\\
{rel.\ dev.\refnote{d}} 
             & 0.97\% & 1.14\% & 1.01\% & 0.87\% & 0.86\% & 0.86\% &
   0.83\% \\
\hline
\end{tabular}
{
\tablenote{a} Using the lowest 38 unoccupied orbitals of s, p and d
symmetry, respectively.\\ 
\tablenote{b} Experimental values from Ref.\
\protect\cite{BashkinStoner:75}.\\
\tablenote{c} Mean value of the absolute deviations from experiment,
(all states tabulated)\\
\tablenote{d} Same as \tablenote{c}, but excluding the $2s \rightarrow
2p$ transition.}
\end{table}

\begin{table}[t]
\caption{\label{table:10}
  Triplet excitation energies for the Be atom,
  calculated from the exact XC potential by
  using approximate XC kernels (in atomic units)}
\begin{tabular}{ccccccccc}
\hline
   &  &\multicolumn{2}{c}{\rm ALDA (xc)} 
      &\multicolumn{2}{c}{\rm TDOEP (X-only)} 
      &\multicolumn{2}{c}{\rm TDOEP (SIC)}\\ 
\cline{3-4} \cline{5-6} \cline{7-8}
   $ k \rightarrow j $  
&$\omega_{jk}$
& SPA    & full\refnote a   
& SPA    & full\refnote a  
& SPA    & full\refnote a  
& Expt. \refnote b\\ 
\hline
{$2s\rightarrow 2p$}& 0.1327 & 0.0982 & 0.0902 & 0.0679 & 0.0000 & 0.0916 &
0.0795 & 0.1002 \\
{$2s\rightarrow 3s$}& 0.2444 & 0.2390 & 0.2387 & 0.2349 & 0.2338 & 0.2431 &
0.2430 & 0.2373 \\
{$2s\rightarrow 3p$}& 0.2694 & 0.2651 & 0.2651 & 0.2647 & 0.2652 & 0.2700 &
0.2705 & 0.2679 \\
{$2s\rightarrow 3d$}& 0.2833 & 0.2807 & 0.2805 & 0.2814 & 0.2813 & 0.2867 &
0.2865 & 0.2827 \\
{$2s\rightarrow 4s$}& 0.2959 & 0.2943 & 0.2943 & 0.2932 & 0.2934 & 0.2953 &
0.2953 & 0.2939 \\
{$2s\rightarrow 4p$}& 0.3046 & 0.3031 & 0.3032 & 0.3031 & 0.3034 & 0.3048 &
0.3049 & 0.3005 \\
{$2s\rightarrow 4d$}& 0.3098 & 0.3087 & 0.3087 & 0.3098 & 0.3089 & 0.3106 &
0.3107 & 0.3096 \\
{$2s\rightarrow 5s$}& 0.3153 & 0.3146 & 0.3146 & 0.3142 & 0.3143 & 0.3150 &
0.3150 & 0.3144 \\
{$2s\rightarrow 5p$}& 0.3193 & 0.3186 & 0.3187 & 0.3187 & 0.3188 & 0.3194 &
0.3194 & 0.3193 \\
{$2s\rightarrow 6s$}& 0.3247 & 0.3243 & 0.3243 & 0.3241 & 0.3242 & 0.3245 &
0.3245 & 0.3242 \\
{$2s\rightarrow 6p$}& 0.3269 & 0.3265 & 0.3265 & 0.3265 & 0.3266 & 0.3269 &
0.3269 & 0.3268 \\
 \hline
{Mean abs. dev.\refnote{c}}  
&  0.0045 &  0.0012 &  0.0020 &  0.0040 &  0.0102 &  0.0026 &  0.0037 \\
{Mean rel.\ dev.}
& 3.53\% & 0.56\% & 1.28\% & 3.31\% & 9.51\% & 1.44\% & 2.55\% \\
\hline
abs.\ dev.\refnote{d}
  & 0.0017 & 0.0012 & 0.0012 & 0.0012 & 0.0013 & 0.0020 & 0.0020 \\
rel.\ dev.\refnote{d}
  &  0.63\% &  0.42\% &  0.41\% & 0.42\% &  0.46\% &  0.72\% &  0.74\%\\
\hline
\end{tabular}
{
\tablenote{a} Using the lowest 38 unoccupied orbitals of s, p and d
symmetry, respectively.\\ 
\tablenote{b} Experimental values from Ref.\
\protect\cite{BashkinStoner:75}.\\
\tablenote{c} Mean value of the absolute deviations from experiment\\
\tablenote{d} Same as \tablenote{c}, but excluding the $2s \rightarrow
2p$ transition.}
\end{table}


For the singlet excitation spectrum, given in table
\ref{table:9}, the TDDFT corrections yield
significantly improved excitation energies compared to 
spectrum of the bare Kohn-Sham eigenvalue differences,
with average errors reduced by a factor of about 3 regardless
of which kernel is used.
The most distinct improvement towards experiment
is achieved for the singlet 2P excitation, where the Kohn-Sham eigenvalue
difference is off by 32\% (61 mHartree) from the experimental value.

\begin{figure}[ht!]
\unitlength1.0cm
\begin{picture}(18,12)
\put(-16.8,10.9){\makebox(18,12){
\includegraphics{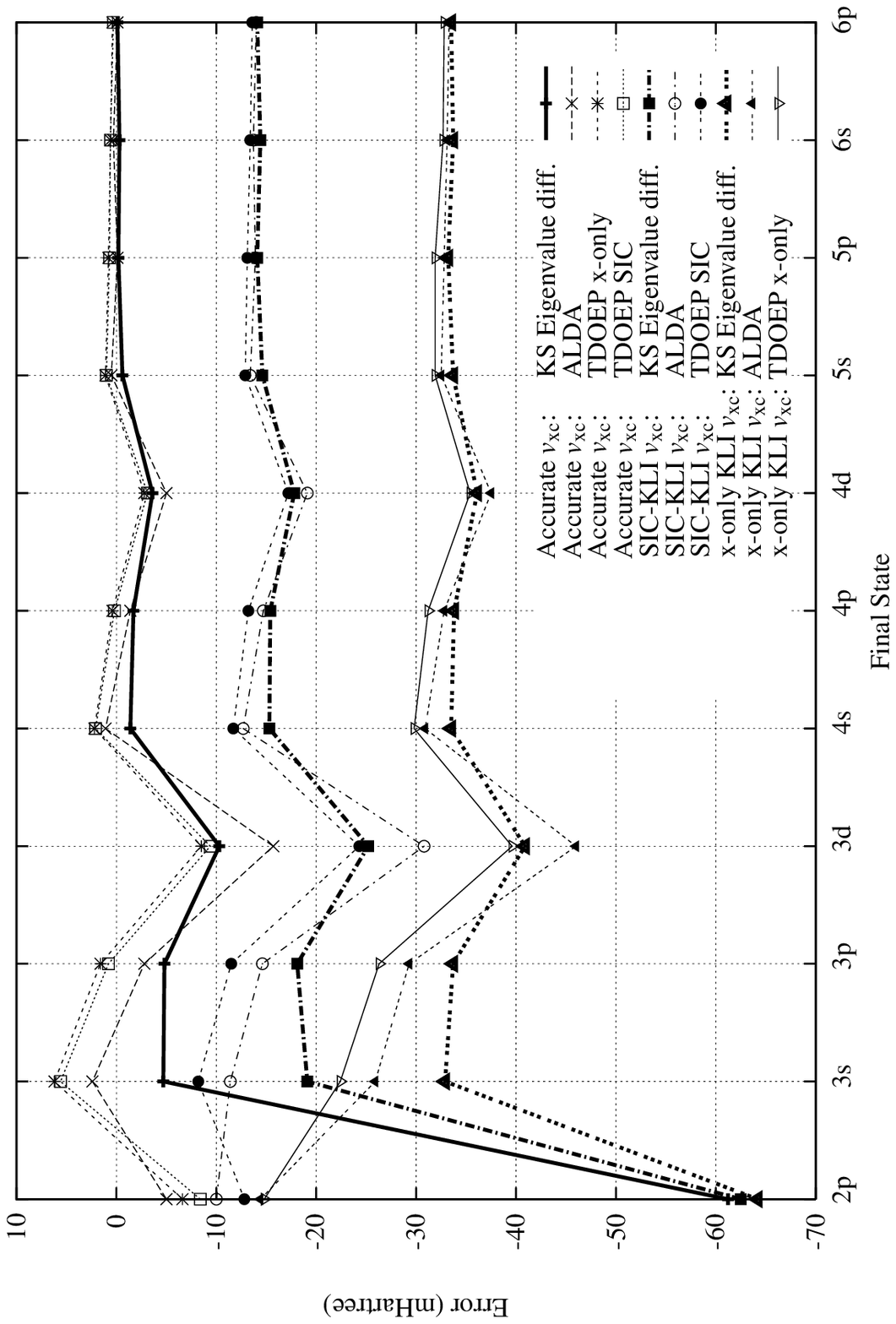}
}}
\end{picture}
\caption{\small\label{fig:3} 
Errors of singlet excitation energies from the ground state of Be,
calculated from the accurate, the OEP-SIC and X-only KLI exchange
correlation potential and with different
approximations for the exchange-correlation kernel (see text).
The errors are given in mHartrees. 
To guide the eye, the errors of the discrete
excitation energies were connected with lines.
}
\end{figure}

For the remaining singlet excitations, 
the TDOEP-SIC kernel yields the best improvement upon the bare 
Kohn-Sham spectrum.
From figure \ref{fig:3}, where the errors for each singlet
excitation energy are plotted, we see two competing effects:
the errors increase with progressing angular momentum
(with the error of the 3d-states being largest), but decrease with 
progressing principal quantum number $n$.   Note that ALDA has the
largest errors for the d-states, presumably due to its inability
to account for orbital nodes.

%
For the triplet spectrum given in table \ref{table:10},
the transition to the 2p state is clearly problematic,
presumably because of its small magnitude.  In particular,
the TDOEP X-only calculation greatly underestimates the downshift
away from the KS eigenvalue difference.  Because of this effect,
we also report average errors with this transition excluded.
All Kohn-Sham
orbital excitations experience a downshift in the 
ALDA and TDOEP X-only calculation.
In ALDA, this leads to an overall improvement of the 
spectrum by a more than a factor of 2.
The downshift in TDOEP X-only results is too strong, and
this behavior is partly corrected in the TDOEP-SIC.
However, due to overcorrections for the higher excitation
energies, the average reduction in error over the Kohn-Sham excitation
spectrum is only a factor of 1.2.

\begin{figure}[ht!]
\unitlength1.0cm
\begin{picture}(18,12)
\put(-16.8,10.9){\makebox(18,12){
\includegraphics{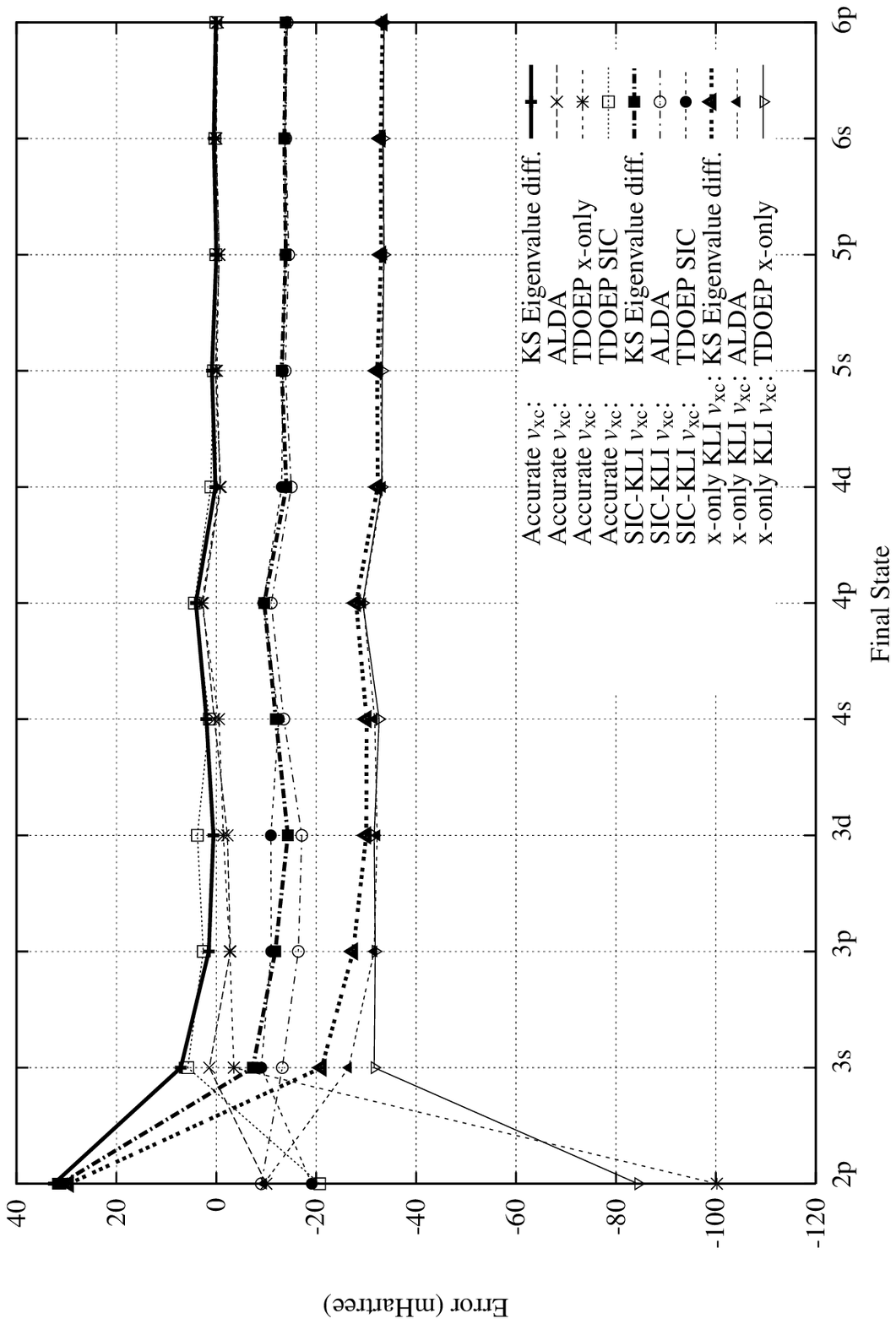}
}}
\end{picture}
\caption{\small\label{fig:4} 
 Errors of triplet excitation energies from the ground state of Be,
 calculated from the accurate, OEP-SIC and X-only KLI exchange
 correlation potential and with different
 approximations for the exchange-correlation kernel (see text).
 The errors are given in mHartrees. 
 To guide the eye, the errors of the discrete
 excitation energies were connected with lines.
}
\end{figure}

The errors for the triplet excitation energies 
are plotted in figure \ref{fig:4}. 
Clearly, the errors of both the Kohn-Sham eigenvalue spectrum and the 
corresponding corrections decrease again with progressing quantum number.
Together with the errors plotted in figure \ref{fig:3} this signals
that the Rydberg-like
transitions to states with high principal quantum number $n$ are 
already close to the eigenvalue differences of the accurate
Kohn-Sham potential. 

\begin{table}
\caption{\label{table:11}
  Singlet-triplet separations in beryllium
  calculated from the exact XC potential by
  using various approximate XC kernels. 
  Calculated from Eq.\ (\protect\ref{eigenwertgl}), 
  using the lowest 38
  unoccupied orbitals of s, p and d symmetry.
  All values are in mHartrees.}
\begin{tabular}{crrrr}
\hline
 &\multicolumn{1}{c}{\rm ALDA} 
 &\multicolumn{2}{c}{\rm TDOEP} \\ \cline{2-4} 
State & XC\refnote a
      & X-only          & SIC    
      & Expt.\refnote b\\ 
\hline
2$P$& 98.7 & 187.3 & 106.0 & 93.8 \\
3$S$& 12.8 &  21.5 &  11.8 & 11.8 \\
3$P$& 6.2  &  10.5 &   4.6 &  6.4 \\
3$D$&-2.6  &   3.8 &  -2.3 & 10.8 \\
4$S$& 4.1  &   6.1 &   4.1 &  3.4 \\
4$P$& 1.7  &   3.4 &   1.6 &  5.8 \\
4$D$&-0.2  &   1.7 &  -0.3 &  3.8 \\
5$S$& 1.8  &   2.7 &   1.9 &  2.4 \\
5$P$& 0.7  &   1.5 &   0.8 &  0.2 \\
6$S$& 1.0  &   1.4 &   1.0 &  0.6 \\
6$P$& 0.3  &   0.8 &   0.4 &  --  \\
\hline
{dev.\refnote c}& 3.0 & 12.4 & 3.8 \\
{rel.\ dev.}    & 68\% &128\% & 75\%\\
\hline\\
\end{tabular} 
\newline \tablenote{a} Including correlation contributions in the form of
Vosko, Wilk and Nusair {\protect\cite{VoskoWilkNusair:80}}.\\
\tablenote{b} Taken from Ref.\ {\protect\cite{BashkinStoner:75}}\\
\tablenote{c} Mean absolute deviation from experiment, excluding the 6$P$
state.
\end{table}

The singlet-triplet separations
from equation (\ref{eigenwertgl})  are given in table 
\ref{table:11} 
for the three different approximate XC kernels. 
Like in helium, the singlet-triplet splittings are overestimated by
about a factor of two for the $S$ and $P$ transitions if the
(diagonal) TDOEP x-only kernel is used. The splittings of the $D$
levels, however, appear too {\em small} by about a factor of two.
By the inclusion of correlation contributions to the kernels, the 
splittings of the $S$ and $P$ levels are consistently (and usually
correctly) reduced.
However, for the D states, this correction is always too large,
and leads to a reversal of the singlet and
triplet energies.%
\footnote{This effect can also be observed in the Helium
  atom. The exact values of the
  singlet-triplet splittings of the $D$ states in helium 
  however, are by two orders of magnitude
  smaller than in beryllium.} 
From the singlet-triplet splitting in Eq.\ (\ref{levels})
which, in the SPA, hold for 
any system since the diagonal terms of $f_{{\rm xc} \sigma \sigma'}$
cancel, this behavior can be traced back to the 
overestimation of correlation contributions in LDA (in small systems).
Self-interaction corrections are not expected to
cure this shortcoming, for the reason that to leading order 
 the self-interaction correction terms cancel  
in the expressions for the splittings, similar to the way   
shown in section
\ref{section:Results_for_the_Helium_Atom}. Accordingly, the
separations in TDOEP-SIC and ALDA are of similar quality, which can be seen
from columns one and three in table \ref{table:11}.
The TDOEP X-only results on the other hand, 
although too small, show the correct ordering
of singlet and triplet levels. 

With increasing excitation energy, the difference
between the results in SPA and the full solution
is reduced,
as was already observed in the case of helium.
The drastic change of the triplet 2$P$ state in TDOEP X-only 
seems to be an artifact of the specific approximation
to the exchange-correlation kernel, since the results
in the SPA and the full calculation 
for this particular excitation energy only differ by  10\% if the 
ALDA is used for $f_{{\rm xc} \sigma \sigma'}$.
 
\subsection{Approximate Kohn-Sham potentials}

The results from using different approximate exchange-correlation 
{\em potentials} for the Be atom
to calculate the Kohn-Sham eigenvalues and
orbitals are given in tables \ref{table:12} to \ref{table:15}.

\begin{table}[t]
\caption{\label{table:12}
  Singlet excitation energies for the Be atom,
  calculated from the X-only KLI potential by
  using approximate XC kernels (in atomic units)}
\begin{tabular}{ccccccc}
\hline 
   &  &\multicolumn{2}{c}{\rm ALDA (xc)} 
      &\multicolumn{2}{c}{\rm TDOEP (X-only)} \\
\cline{3-4} \cline{5-6} 
  $ k \rightarrow j $  
&$\omega_{jk}$
& SPA    & full\refnote a   
& SPA    & full\refnote a  
& Expt. \refnote b\\ 
\hline
{$2s\rightarrow 2p$}& 0.1297 & 0.1990 & 0.1795 & 0.1958 & 0.1791 & 0.1939 \\
{$2s\rightarrow 3s$}& 0.2162 & 0.2245 & 0.2232 & 0.2288 & 0.2267 & 0.2491 \\
{$2s\rightarrow 3p$}& 0.2405 & 0.2415 & 0.2449 & 0.2465 & 0.2479 & 0.2742 \\
{$2s\rightarrow 3d$}& 0.2527 & 0.2480 & 0.2476 & 0.2541 & 0.2540 & 0.2936 \\
{$2s\rightarrow 4s$}& 0.2638 & 0.2663 & 0.2664 & 0.2674 & 0.2675 & 0.2973 \\
{$2s\rightarrow 4p$}& 0.2725 & 0.2727 & 0.2735 & 0.2745 & 0.2751 & 0.3063 \\
{$2s\rightarrow 4d$}& 0.2773 & 0.2758 & 0.2759 & 0.2780 & 0.2780 & 0.3134 \\
{$2s\rightarrow 5s$}& 0.2822 & 0.2833 & 0.2834 & 0.2838 & 0.2840 & 0.3159 \\
{$2s\rightarrow 5p$}& 0.2863 & 0.2864 & 0.2867 & 0.2872 & 0.2876 & 0.3195 \\
{$2s\rightarrow 6s$}& 0.2913 & 0.2918 & 0.2919 & 0.2921 & 0.2923 & 0.325  \\
{$2s\rightarrow 6p$}& 0.2935 & 0.2936 & 0.2937 & 0.2940 & 0.2942 & 0.327  \\
 \hline
{Mean abs. dev.\refnote{c}}
                    & 0.0372 & 0.0311 & 0.0317 & 0.0288 & 0.0299 \\
{Mean rel.\ dev.\refnote{d}}              
                    &13.5\% &10.4\% &10.7\% & 9.5\% &10.1\% \\
\hline
{abs.\ dev.\refnote{d}}
 &  0.0345 & 0.0337 &  0.0334 &  0.0315 &  0.0314 \\
{rel.\ dev.\refnote{d}}
 &11.49\% &11.18\% &11.07\% &10.39\% &10.38\% \\
\hline
\end{tabular}
\newline%
{\tablenote{a} Using the lowest 38 unoccupied orbitals of s, p and d
symmetry, respectively.\\ 
\tablenote{b} Experimental values from Ref.\
\protect\cite{BashkinStoner:75}.\\
\tablenote{c} Mean value of the absolute deviations from experiment\\
\tablenote{d} Same as \tablenote{c}, but excluding the $2s \rightarrow
2p$ transition.}
\end{table}

\begin{table}[t]
\caption{\label{table:13}
  Triplet excitation energies for the Be atom,
  calculated from the X-only KLI potential by
  using approximate XC kernels (in atomic units)}
\begin{tabular}{ccccccc}
\hline
   &  &\multicolumn{2}{c}{\rm ALDA (xc)} 
      &\multicolumn{2}{c}{\rm TDOEP (X-only)} \\
\cline{3-4} \cline{5-6} 
   $ k \rightarrow j $  
&$\omega_{jk}$
& SPA    & full\refnote a   
& SPA    & full\refnote a  
& Expt. \refnote b\\ 
\hline
{$2s\rightarrow 2p$}& 0.1297 & 0.0980 & 0.0907 & 0.0692 & 0.0158 & 0.1002 \\
{$2s\rightarrow 3s$}& 0.2162 & 0.2112 & 0.2108 & 0.2069 & 0.2057 & 0.2373 \\
{$2s\rightarrow 3p$}& 0.2405 & 0.2362 & 0.2363 & 0.2353 & 0.2361 & 0.2679 \\
{$2s\rightarrow 3d$}& 0.2527 & 0.2506 & 0.2505 & 0.2512 & 0.2511 & 0.2827 \\
{$2s\rightarrow 4s$}& 0.2638 & 0.2622 & 0.2622 & 0.2611 & 0.2613 & 0.2939 \\
{$2s\rightarrow 4p$}& 0.2725 & 0.2710 & 0.2711 & 0.2709 & 0.2712 & 0.3005 \\
{$2s\rightarrow 4d$}& 0.2773 & 0.2763 & 0.2763 & 0.2765 & 0.2765 & 0.3096 \\
{$2s\rightarrow 5s$}& 0.2822 & 0.2815 & 0.2816 & 0.2811 & 0.2812 & 0.3144 \\
{$2s\rightarrow 5p$}& 0.2863 & 0.2856 & 0.2857 & 0.2856 & 0.2857 & 0.3193 \\
{$2s\rightarrow 6s$}& 0.2913 & 0.2909 & 0.2909 & 0.2907 & 0.2908 & 0.3242 \\
{$2s\rightarrow 6p$}& 0.2935 & 0.2931 & 0.2932 & 0.2931 & 0.2932 & 0.3268 \\
 \hline
{Mean abs. dev.\refnote{c}}  
 & 0.0300 & 0.0291 &  0.0298 &  0.0323 &  0.0371 \\
{Mean rel.\ dev.}
&11.8\% & 9.9\% &10.6\% &12.8\% &17.6\% \\
\hline
abs.\ dev.\refnote{d}
&  0.0300 &  0.0318 &  0.0318 &  0.0324 &  0.0324 \\
rel.\ dev.\refnote{d}
&10.1\% &10.7\% &10.7\% &11.0\% &11.0\% \\
\hline
\end{tabular}
\newline%
{\tablenote{a} Using the lowest 38 unoccupied orbitals of s, p and d
symmetry, respectively.\\ 
\tablenote{b} Experimental values from Ref.\
\protect\cite{BashkinStoner:75}.\\
\tablenote{c} Mean value of the absolute deviations from experiment\\
\tablenote{d} Same as \tablenote{c}, but excluding the $2s \rightarrow
2p$ transition.}
\end{table}

\begin{table}[t]
\caption{\label{table:14}
  Singlet excitation energies for the Be atom,
  calculated from the SIC-KLI potential by
  using approximate XC kernels (in atomic units)}
\begin{tabular}{ccccccc}
\hline 
   &  &\multicolumn{2}{c}{\rm ALDA (xc)} 
      &\multicolumn{2}{c}{\rm TDOEP (SIC)} \\
\cline{3-4} \cline{5-6} 
  $ k \rightarrow j $  
&$\omega_{jk}$
& SPA    & full\refnote a   
& SPA    & full\refnote a  
& Expt. \refnote b\\ 
\hline
{$2s\rightarrow 2p$}& 0.1314 & 0.2030 & 0.1839 & 0.1968 & 0.1811 & 0.1939 \\
{$2s\rightarrow 3s$}& 0.2300 & 0.2390 & 0.2377 & 0.2429 & 0.2409 & 0.2491 \\
{$2s\rightarrow 3p$}& 0.2561 & 0.2566 & 0.2596 & 0.2614 & 0.2627 & 0.2742 \\
{$2s\rightarrow 3d$}& 0.2684 & 0.2632 & 0.2628 & 0.2694 & 0.2693 & 0.2936 \\
{$2s\rightarrow 4s$}& 0.2820 & 0.2844 & 0.2846 & 0.2855 & 0.2856 & 0.2973 \\
{$2s\rightarrow 4p$}& 0.2909 & 0.2910 & 0.2916 & 0.2926 & 0.2931 & 0.3063 \\
{$2s\rightarrow 4d$}& 0.2956 & 0.2943 & 0.2943 & 0.2961 & 0.2962 & 0.3134 \\
{$2s\rightarrow 5s$}& 0.3013 & 0.3023 & 0.3025 & 0.3028 & 0.3030 & 0.3159 \\
{$2s\rightarrow 5p$}& 0.3054 & 0.3054 & 0.3056 & 0.3062 & 0.3064 & 0.3195 \\
{$2s\rightarrow 6s$}& 0.3106 & 0.3112 & 0.3113 & 0.3114 & 0.3116 & 0.325  \\
{$2s\rightarrow 6p$}& 0.3129 & 0.3129 & 0.3130 & 0.3133 & 0.3134 & 0.327  \\
 \hline
{Mean abs. dev.\refnote{c}}
&  0.0210 &  0.0155 &  0.0153 &  0.0130 &  0.0138 \\
{Mean rel.\ dev.\refnote{d}}              
& 8.07\% & 5.29\% & 5.25\% & 4.32\% & 4.78\% \\
\hline
abs.\ dev.\refnote{d}
& 0.0168 & 0.0161 & 0.0158 & 0.0140 & 0.0139 \\
rel.\ dev.\refnote{d}
& 5.65\% & 5.35\% & 5.26\% & 4.60\% & 4.60\% \\
\hline
\end{tabular}
\newline%
{\tablenote{a} Using the lowest 38 unoccupied orbitals of s, p and d
symmetry, respectively.\\ 
\tablenote{b} Experimental values from Ref.\
\protect\cite{BashkinStoner:75}.\\
\tablenote{c} Mean value of the absolute deviations from experiment\\
\tablenote{d} Same as \tablenote{c}, but excluding the $2s \rightarrow
2p$ transition.}
\end{table}

\begin{table}[t]
\caption{\label{table:15}
  Triplet excitation energies for the Be atom,
  calculated from the SIC-KLI potential by
  using approximate XC kernels (in atomic units)}
\begin{tabular}{ccccccc}
\hline
   &  &\multicolumn{2}{c}{\rm ALDA (xc)} 
      &\multicolumn{2}{c}{\rm TDOEP (SIC)} \\
\cline{3-4} \cline{5-6} 
   $ k \rightarrow j $  
&$\omega_{jk}$
& SPA    & full\refnote a   
& SPA    & full\refnote a  
& Expt. \refnote b\\ 
\hline
{$2s\rightarrow 2p$}& 0.1314 & 0.0987 & 0.0912 & 0.0925 & 0.0811 & 0.1002 \\
{$2s\rightarrow 3s$}& 0.2300 & 0.2245 & 0.2241 & 0.2284 & 0.2283 & 0.2373 \\
{$2s\rightarrow 3p$}& 0.2561 & 0.2515 & 0.2515 & 0.2563 & 0.2569 & 0.2679 \\
{$2s\rightarrow 3d$}& 0.2684 & 0.2658 & 0.2656 & 0.2720 & 0.2718 & 0.2827 \\
{$2s\rightarrow 4s$}& 0.2820 & 0.2804 & 0.2804 & 0.2814 & 0.2814 & 0.2939 \\
{$2s\rightarrow 4p$}& 0.2909 & 0.2894 & 0.2895 & 0.2910 & 0.2911 & 0.3005 \\
{$2s\rightarrow 4d$}& 0.2956 & 0.2946 & 0.2946 & 0.2965 & 0.2965 & 0.3096 \\
{$2s\rightarrow 5s$}& 0.3013 & 0.3006 & 0.3006 & 0.3010 & 0.3011 & 0.3144 \\
{$2s\rightarrow 5p$}& 0.3054 & 0.3047 & 0.3048 & 0.3055 & 0.3055 & 0.3193 \\
{$2s\rightarrow 6s$}& 0.3106 & 0.3103 & 0.3103 & 0.3105 & 0.3105 & 0.3242 \\
{$2s\rightarrow 6p$}& 0.3129 & 0.3125 & 0.3126 & 0.3129 & 0.3129 & 0.3268 \\
 \hline
{Mean abs. dev.\refnote{c}}  
& 0.0141 & 0.0131 & 0.0138 & 0.0117 & 0.0127 \\
{Mean rel.\ dev.}
& 6.58\% & 4.52\% & 5.22\% & 4.39\% & 5.41\% \\
\hline
abs.\ dev.\refnote{d}
& 0.0123 &  0.0142 &  0.0143 &  0.0121 &  0.0121 \\
rel.\ dev.\refnote{d}
& 4.13\% & 4.83\% & 4.84\% & 4.06\% & 4.04\% \\
\hline
\end{tabular}
\newline%
{\tablenote{a} Using the lowest 38 unoccupied orbitals of s, p and d
symmetry, respectively.\\ 
\tablenote{b} Experimental values from Ref.\
\protect\cite{BashkinStoner:75}.\\
\tablenote{c} Mean value of the absolute deviations from experiment\\
\tablenote{d} Same as \tablenote{c}, but excluding the $2s \rightarrow
2p$ transition.}
\end{table}

The errors towards the experimental excitation energies are compiled
in figures \ref{fig:3} and \ref{fig:4} for the singlet and triplet
series. 
Looking first at the spectra of the bare Kohn-Sham eigenvalues
(represented in figures \ref{fig:3} and \ref{fig:4} by the points 
connected with thick lines),
we notice that the ``HOMO-LUMO'' gap is almost independent
of the approximation of $v_{\rm xc}$ employed.  This is in sharp
contrast with the He atom case.
The correlation contributions cancel for the lowest excitation energy,
and we must classify this as a non-Rydberg state.  
For the higher states, the situation is different:
Starting from the excitation
to the 3s level, the series of single-particle  energy differences 
appear almost {\em uniformly shifted} with respect to the series of the
exact
potential, preserving the typical pattern of their deviation from the
experimentally measured spectrum.
The shifts amount to -14 mH for the OEP-SIC
potential, and -34 mH for the X-only KLI potential.
As in helium,
these shifts are equal to the 
differences in the eigenvalues of the highest occupied Kohn-Sham
orbital: For the accurate potential 
($\epsilon_{2s}^{\rm accurate}=-0.3426$ a.u.\
\cite{FilippiGonzeUmrigar:96}), the highest occupied orbitals are more
strongly bound than in OEP-SIC 
($\epsilon_{2s}^{\rm OEP-SIC}=-0.3285$ a.u.) and in X-only KLI
($\epsilon_{2s}^{\rm X-only KLI}=-0.3089$ a.u.).
Thus, among the virtual states, only the 2p orbital is 
appreciably influenced by the 
details of the ground state potential. For the higher lying states, the
long-range behavior of the Kohn-Sham potential dominates. Its 
$-1/r$ behavior is correctly reproduced both in X-only KLI as well as in
OEP-SIC.   For larger systems, more low-lying excitations can
be accurately approximated, but eventually, for any finite system,
the Rydberg excitations
will show errors due to errors in the ionization potential.
Casida et al.\cite{CJCS98} have studied which excitations
can be well-approximated  with present functional approximations
to the potential.

Regarding the corrections for the singlet excitation energies
calculated from Eq.\ (\ref{eigenwertgl}),
the first
excited state (2p) experiences the largest correction, irrespective
of the exchange-correlation potential employed. Moreover, the
results using different approximate exchange-correlation kernels agree
within 10 mH.
For the remaining singlet excitation energies, 
the calculated corrections using the approximate Kohn Sham orbitals
are almost identical to the corrections which are obtained from using  
the accurate Kohn-Sham orbitals.
Hence, in figure \ref{fig:3} the errors for the excitations to 3s
through 6p show the same pattern of deviations, only
shifted by the error in the respective eigenvalue of
the 2s orbital. On average, 
the resulting singlet excitation energies are closest to experiment,
if the approximate exchange-correlation potential $v_{\rm xc}$ is
combined with the corresponding approximation of the
exchange-correlation kernel $f_{{\rm xc}\, \sigma \sigma'}$.

From tables \ref{table:13} and \ref{table:15} as well
as from figure \ref{fig:4},
the behavior of the triplet spectra is similar, but
less unequivocal for the triplet 2p state.
For this particular state, the corrections spread on the order of
100mH, prevalently due to the significant overcorrection of the 
X-only TDOEP kernels.
However, the resulting triplet 2p excitation energy almost exclusively
depends on the approximation to the  exchange correlation kernel
rather than on
the exchange-correlation potential employed.
On the average,
apart from the higher excitations in OEP-SIC (c.f.\ table \ref{table:14}),
the best triplet 
spectra are obtained if the ALDA is used for the exchange-correlation
kernel, but this appears to be a fortuitous cancellation of errors.
The approximate Kohn-Sham excitation energies, except for the 2p state,
are already incorrectly lower than the experimental triplet levels. Any further
lowering, although correct for the eigenvalue-differences of the
exact Kohn-Sham potential, actually worsens the triplet spectra which
are calculated on the basis of an approximate exchange-correlation
potential. Since the shifts are reduced by 
correlation contributions in the kernels,
the over-corrections become less severe for the ALDA and TDOEP-SIC kernels.
Another apparent error cancellation is
that when calculating the lowest excitation energy 
($2s\rightarrow 2p$) from approximate exchange-correlation potentials,
the SPA-results are always closer to experiment than the full
results.  This might be related to the fact that for TDOEP X-only,
SPA yields the exact first-order shift in energy levels in G{\"o}rling-Levy
perturbation theory\cite{GS99}, while the ``full'' calculation does 
not\cite{BPG00}.   In cases where there are large differences between
SPA and full results, the SPA might be more reliable for these reasons.

The fact that the corrections to the Kohn-Sham eigenvalue differences 
only weakly depend on the approximation of the exchange-correlation
potential $v_{\rm xc}$, is also reflected in 
table \ref{table:16}, where the singlet-triplet separations in Be,
calculated
using the X-only KLI and OEP-SIC potentials are given.
The numerical values are close to the results for the accurate 
Be exchange correlation potential
in table \ref{table:11}. 
Again, the obtained splittings 
are more sensitive to the approximation of $f_{{\rm xc} \sigma
  \sigma'}$ than to the approximation of the potential $v_{\rm xc}$.

\begin{table}[t]
\caption{\small\label{table:16}
  Singlet-triplet separations in beryllium
  calculated from the X-only potential 
  and the SIC-potential and by 
  using various approximate XC kernels. 
  Calculated from Equation (\protect\ref{eigenwertgl}), 
  using the lowest 38
  unoccupied orbitals of s, p and d symmetry, respectively.
  All values are in mHartrees.}
\begin{tabular}{crrrrr}
\hline
 &\multicolumn{2}{c} {X-only} &
  \multicolumn{2}{c} {SIC}    &\\ \cline{2-3} \cline{4-5}
State & $f_{\rm xc}^{\rm ALDA}$ & $f_{\rm x}^{\rm TDOEP}$
      & $f_{\rm xc}^{\rm ALDA}$ & $f_{\rm xc}^{\rm TDOEP-SIC}$
      & Expt.\refnote b\\ 
\hline
2$P$& 88.9 & 163.3 & 92.7 & 100.1 &  93.8 \\
3$S$& 12.4 &  21.0 & 13.6 &  12.5 &  11.8 \\
3$P$&  8.6 &  11.8 &  8.0 &   5.9 &   6.4 \\
3$D$& -2.9 &   2.9 & -2.8 &  -2.5 &  10.8 \\
4$S$&  4.1 &   6.2 &  4.2 &   4.2 &   3.4 \\
4$P$&  2.4 &   3.9 &  2.1 &   2.0 &   5.8 \\
4$D$& -0.4 &   1.4 & -0.2 &  -0.4 &   3.8 \\
5$S$&  1.9 &   2.7 &  1.8 &   1.9 &   2.4 \\
5$P$&  1.0 &   1.8 &  0.8 &   0.9 &   0.2 \\
6$S$&  1.0 &   1.5 &  1.0 &   0.4 &   0.6 \\
6$P$&  0.5 &   1.0 &  0.4 &   0.5 &   --  \\
\hline
{dev.\refnote c}& 
     & 3.1 &  10.2 &  2.8 &   3.1 \\
{rel.\ dev.}    &
     & 85\%& 145\% & 75\% & 75\% \\
\hline\\
\end{tabular} 
\newline \tablenote{a} Including correlation contributions in the form of
Vosko, Wilk and Nusair {\protect\cite{VoskoWilkNusair:80}}.\\
\tablenote{b} Taken from Ref.\ {\protect\cite{BashkinStoner:75}}\\
\tablenote{c} Mean absolute deviation from the exact values
\end{table}

\section{Summary and Conclusion}

In this work we aimed at an assessment of the influence of the three
different types of approximations (i.e.\ (i) the XC potential $v_{\rm
xc}$, (ii) the XC kernel $f_{\rm xc}$ and (iii) truncation of the
space of virtual excitations)
inherent in the calculation of excitation energies from TDDFT. 
We calculated the 
discrete optical spectra of helium and beryllium, two of the 
spectroscopically best known elements, using the exact
exchange-correlation potential, the KLI-X-only potential and 
the the KLI-SIC potential for $v_{xc}$
(all three potentials are falling off like $-1/r$ as $r\rightarrow\infty$).
These were combined with 
three approximations for the XC kernel: The adiabatic LDA (ALDA),
the TDOEP X-only kernel and the TDOEP-SIC kernel. The results are
given both in the single-pole approximation (SPA) and for a 
``full'' calculation, where as many virtual states as possible
(typically about 30 to 40) entered the calculation.
The analysis of these combinations reveals the 
following trends:
First of all, the choice of $v_{\rm xc}$ on the calculated spectrum 
has the largest effect on the calculated spectra.
The inaccuracies introduced by {\em approximate} ground state
Kohn-Sham potentials (even those including exact exchange)
can be quite substantial.
This is especially true for the higher excited states,
which appear almost uniformly shifted from the true 
excitation energies.
We observe that this shift is closely related 
to the absolute value of the highest occupied eigenvalue,
which, in exact DFT, is equal to the first ionization
potential of the system at hand.

For the lower excitation energies, an error cancellation
occurs, making these excitations less sensitive to the 
choice of the exchange-correlation potential.
This error cancellation however, ceases to work
the more the excited states behave like Rydberg states.
For Helium, this is already the case for the first excited state.
Hence, in improving the calculation of excitation energies
from TDDFT requires an improved exchange-correlation potential
in the first place. The most important requirement for such
a potential would be that its highest occupied eigenvalue
reproduces the experimental ionization potential as closely
as possible. 
Empirically, one could introduce a ``scissors-operator'' 
similar to the one introduced by Levine and Allan \cite{LevinaAllan:89}, 
which shifts the Rydberg states by a constant being equal to the 
difference between the highest occupied eigenvalue and the 
negative of the experimental ionization potential. But
such a  procedure would not produce a
{\em first-principles} calculation.
In our opinion however, the construction of
approximate exchange-correlation
potentials based on orbital functionals would be the method
of choice for the future.
 
The effect of the choice of the exchange-correlation {\em kernel} on
the calculated spectra, in turn, is much less pronounced.
However, its relative importance increases
whenever the ``first-order''
effects, originating from $v_{\rm xc}$ cancel.
This is the case for the values of the
singlet-triplet splittings
and the lower excitation energies of Be.
For these ``second-order effects'', the correlation 
contributions contained in $f_{\rm xc}$ are important.
We observe that the ALDA for the XC kernels
already leads to quite reasonable results which are
only marginally improved by using the more complicated
TDOEP-SIC kernel. Besides the missing frequency dependence,
correlation contributions 
are hard to model on top of an exact exchange treatment and one might
speculate that the ALDA takes advantage from a fortuitous error
cancellation between exchange and correlation effects.
Again, we expect only orbital functionals to manage
a marked improvement over the ALDA, which, up to now, can
has been the workhorse of TDDFT.

Finally, the inevitable truncation of the space of virtual excitations 
is appreciable only for the lowest lying states.
In most cases, the results of the 
single-pole approximation (SPA), which, in the
nondegenrate case, merely requires a pair of ``initial'' and ``final'' 
KS states are close to the results obtained from 
using more configurations.

We thank C.\ Umrigar for providing us with the data of the exact
exchange-correlation potentials for helium and beryllium. This work was
supported in part by the Deutsche Forschungsgemeinschaft and by NATO.
K.B. acknowledges support of the Petroleum Research Fund and of
NSF grant CHE-9875091.


\begin{thebibliography}{10}

\bibitem{HohenbergKohn:64}
P. Hohenberg and W. Kohn, Phys.\ Rev. {\bf 136},  B864  (1964).

\bibitem{GunnarssonLundqvist:76}
O. Gunnarsson and {B.I.~Lundqvist}, Phys.\ Rev.\ B {\bf 13},  4274  (1976).

\bibitem{ZieglerRaukBaerends:77}
T. Ziegler, A. Rauk, and {E.J.~Baerends}, Theoret.\ Chim.\ Acta {\bf 43},
261
  (1977).

\bibitem{vonBarth:79}
{U.~von Barth}, Phys.\ Rev.\ A {\bf 20},  1693  (1979).

\bibitem{Theophilou:79}
A. Theophilou, J.\ Phys.\ C {\bf 12},  5419  (1979).

\bibitem{Kohn:86}
W. Kohn, Phys.\ Rev.\ A {\bf 34},  5419  (1986).

\bibitem{GrossOliveiraKohn:88a}
{E.K.U.~Gross}, {L.N.~Oliveira}, and W. Kohn, Phys.\ Rev.\ A {\bf 37},  2805
  (1988).

\bibitem{GrossOliveiraKohn:88b}
{E.K.U.~Gross}, {L.N.~Oliveira}, and W. Kohn, Phys.\ Rev.\ A {\bf 37},  2809
  (1988).

\bibitem{OliveiraGrossKohn:88}
{L.N.~Oliveira}, {E.K.U.~Gross}, and W. Kohn, Phys.\ Rev.\ A {\bf 37},  2821
  (1988).

\bibitem{Nagy:90}
A. Nagy, Phys.\ Rev.\ A {\bf 42},  4388  (1990).

\bibitem{Nagy:95}
A. Nagy, Int.\ J.\ Quantum Chem.\ Symp. {\bf 29},  297  (1995).

\bibitem{Levy:95}
M. Levy, Phys.\ Rev.\ A {\bf 52},  R4313  (1995).

\bibitem{Goerling:96}
A. {G\"orling}, Phys.\ Rev.\ A {\bf 54},  3912  (1996).

\bibitem{PetersilkaGossmannGross:96}
M. Petersilka, {U.J.~Gossmann}, and {E.K.U~Gross}, Phys.\ Rev.\ Lett. {\bf
76},
   1212  (1996).

\bibitem{RungeGross:84}
E. Runge and {E.K.U.~Gross}, Phys.\ Rev.\ Lett. {\bf 52},  997  (1984).

\bibitem{PetersilkaGross:96}
M. Petersilka and E.K.U. Gross, Int.\ J.\ Quantum Chem. {\bf 60},  181
(1996).

\bibitem{BauernschmittAhlrichs:96}
R. Bauernschmitt and R. Ahlrichs, Chem.\ Phys.\ Lett. {\bf 256},  454
(1996).

\bibitem{C96}
{\em Time-dependent density functional response theory of molecular
systems:  Theory, computational methods, and functionals}, M.E. Casida, in {\em
Recent developments and applications in density functional theory},
ed. J.M. Seminario (Elsevier, Amsterdam, 1996).


\bibitem{BauernschmittEtAl:97}
R. Bauernschmitt, M. Hauser, O. Treutler, and R. Ahlrichs, Chem.\ Phys.\
Lett.
  {\bf 264},  573  (1997).

\bibitem{Grabo:97}
T. Grabo, Ph.D. thesis, Universit{\"{a}}t W{\"{u}}rzburg, 1997.

\bibitem{GisbergenEtAl:98}
S.J.A. van Gisbergen, F. Kootstra, P.R.T. Schipper, O.V. Gritsenko, J.G.
  Snijders, and E.J. Baerends, Phys.\ Rev.\ A {\bf 57},  2556  (1998).

\bibitem{CCS98}
M.E. Casida, K.C. Casida, and D.R. Salahub, Int. J. Quant. Chem. {\bf 70}, 933 (1998).

\bibitem{CJCS98}
M.E. Casida, C. Jamorski, K.C. Casida, and D.R. Salahub,
J. Chem. Phys. {\bf 108}, 4439 (1998).

\bibitem{GraboPetersilkaGross:99}
T. Grabo, M. Petersilka, and E.K.U. Gross, Phys.\ Rev.\ A  (1999), in press.

\bibitem{Sb99}
D. Sundholm, Chem. Phys. Lett., {\bf 302}, 480 (1999).



\bibitem{PerdewZunger:81}
J.P. Perdew and A. Zunger, Phys.\ Rev.\ B {\bf 23},  5048  (1981).

\bibitem{KriegerLiIafrate:90a}
{J.B.} Krieger, Y. Li, and {G.J.} Iafrate, Phys.\ Lett.\ A {\bf 146},  256
  (1990).

\bibitem{KriegerLiIafrate:92b}
{J.B.} Krieger, Y. Li, and {G.J.} Iafrate, Phys.\ Rev.\ A {\bf 45},  101
  (1992).

\bibitem{KriegerLiIafrate:92c}
{J.B.} Krieger, Y. Li, and {G.J.} Iafrate, Phys.\ Rev.\ A {\bf 46},  5453
  (1992).

\bibitem{KriegerLiIafrate:95}
{J.B.} Krieger, Y. Li, and {G.J.} Iafrate,  in {\em Density Functional
Theory},
  Vol.~337 of {\em NATO ASI Series B}, edited by {E.K.U.} Gross and R.M.
  Dreizler (Plenum Press, New York, 1995), p.\ 191.

\bibitem{ChenEtAl:96}
J.\ Chen, J.B.\ Krieger, Y.\ Li, and G.J.\ Iafrate, Phys.\ Rev.\ A {\bf 54},
  3939  (1996).

\bibitem{FetterWalecka:71}
A.L.\ Fetter and J.D.\ Walecka, {\em Quantum Theory of Many-Particle
Systems}
  (McGraw-Hill, New York, 1971).

\bibitem{LiuVosko:89}
{K.L.~Liu} and {S.H.~Vosko}, Can.\ J.\ Phys. {\bf 67},  1015  (1989).

\bibitem{GrossDobsonPetersilka:96}
E.K.U. Gross, J.F. Dobson, and M. Petersilka,  in {\em Density Functional
Theory
  II}, Vol.~181 of {\em Topics in Current Chemistry}, edited by R.F.
Nalewajski
  (Springer, Berlin, 1996), p.\ 81.

\bibitem{GrossKohn:90}
{E.K.U.~Gross} and W. Kohn, Adv.~Quant.~Chem. {\bf 21},  255  (1990).

\bibitem{GL93}
A. G{\"o}rling and M. Levy, Phys. Rev. B {\bf 47}, 13105 (1993).


\bibitem{SharpHorton:53}
{R.T.} Sharp and {G.K.} Horton, Phys.\ Rev. {\bf 90},  317  (1953).

\bibitem{TalmanShadwick:76}
{J.D.} Talman and {W.F.} Shadwick, Phys.\ Rev.\ A {\bf 14},  36  (1976).

\bibitem
{UllrichGossmannGross:95a}
{C.A.} Ullrich, {U.J.} Gossmann, and {E.K.U.} Gross, Phys.\ Rev.\ Lett. {\bf
  74},  872  (1995).

\bibitem{L98}
R. van Leeuwen, Phys. Rev. Lett. {\bf 80}, 1280 (1998).

\bibitem{PetersilkaGossmannGross:98}
M. Petersilka, U. Gossmann, and E.K.U. Gross,  in {\em Electronic Density
  Functional Theory: Recent Progress and New Directions}, edited by
G.~Vignale
  J.F.~Dobson and M.P. Das (Plenum, New York, 1998), p.\ 177.

\bibitem{GKKG98}
{\em Orbital functionals in density functional theory: the optimized effective
potential method}, T. Grabo, T. Kreibich, S. Kurth, and E.K.U. Gross, in {\em
Strong Coulomb correlations in electronic structure:  Beyond the local density
approximation}, ed. V.I. Anisimov (Gordon and Breach, Tokyo, 1998).


\bibitem{GDP96}
{\em Density Functional Theory of Time-Dependent Phenomena},
E.K.U.~Gross, J.F.~Dobson, and M.~Petersilka, Topics in Current Chemisty, {\bf 181}, 81 (1996).

 
\bibitem{ChernyshevaCherepkovRadojevic:76}
L.V.\ Chernysheva, N.A.\ Cherepkov, and V.\ Radojevi{\'c}, Comput.\ Phys.\
  Commun. {\bf 11},  57  (1976).

\bibitem{UmrigarGonze:94}
C.J.\ Umrigar and X.\ Gonze, Phys.\ Rev.\ A {\bf 50},  3827  (1994).

\bibitem{KonoHattori:84}
A.\ Kono and S.\ Hattori, Phys.\ Rev.\ A {\bf 29},  2981  (1984).

\bibitem{SUG98}
{\em Relationship of Kohn-Sham eigenvalues to excitation energies},
A. Savin, C.J. Umrigar, X. Gonze, Chem. Phys. Lett. {\bf 288}, 391 (1998).

 
\bibitem{BPG00}
{\em A hybrid functional for the exchange-correlation kernel in time-dependent density functional theory},
K. Burke, M. Petersilka, and E.K.U. Gross, in {\em Recent advances in density functional methods, vol. III},
ed. P. Fantucci and A. Bencini (World Scientific Press, 2000).

 
\bibitem{VoskoWilkNusair:80}
S.H. Vosko, L. Wilk, and M. Nusair, Can.\ J.\ Phys. {\bf 58},  1200  (1980).

\bibitem{EngelChevaryMacdonaldVosko:92}
E.\ Engel, {J.A.}\ Chevary, {L.D.}\ Macdonald, and {S.H.}\ Vosko, Z.\ Phys.\
D
  {\bf 23},  7  (1992).

\bibitem{AlmbladhvonBarth:85}
C.O.\ Almbladh and U.\ von Barth, Phys.\ Rev.\ B {\bf 31},  3231  (1985).

\bibitem{Zuelicke:85}
L.\ Z\"ulicke, {\em Quantenchemie} (H\"uthig, Heidelberg, 1985), Vol.~2.

\bibitem{GrahamEtAl:86}
R.L.\ Graham, D.L.\ Yeager, J.\ Olsen, P.\ J{\o}rgensen, R.\ Harrison, S.\
  Zarrabian, and R.\ Bartlett, J.\ Chem.\ Phys. {\bf 85},  6544  (1986).

\bibitem{UmrigarGonze:93}
C.J.\ Umrigar and X.\ Gonze, a preliminary version was published in {\it
High
  Performance Computing and its Application to the Physical Sciences},
  proceedings of the Mardi Gras '93 Conference, edited by D. A. Browne et
al.,
  (World Scientific, Singapore, 1993) (unpublished).

\bibitem{BashkinStoner:75}
S.\ Bashkin and J.A.\ {Stoner, Jr.}, {\em Atomic Energy Levels and Grotrian
  Diagrams I} (North-Holland, Amsterdam, 1975).

\bibitem{FilippiGonzeUmrigar:96}
C.\ Filippi, X.\ Gonze, and C.J.\ Umrigar,  in {\em Recent Developments and
  Applications of Density Functional Theory}, edited by J.M.\ Seminario
  (Elsevier, Amsterdam, 1996).

\bibitem{GS99}
X. Gonze and M. Scheffler, Phys. Rev.
Lett. {\bf 82}, 4416 (1999).
 
\bibitem{LevinaAllan:89}
Z.H. Levine and D.C. Allan, Phys.\ Rev.\ Lett. {\bf 63},  1719  (1989).

\end{thebibliography}
\end{document}